\documentclass[preprints,review,accept,pdftex,moreauthors,a4paper]{Definitions/mdpi}

\firstpage{1} 
\pubvolume{1}
\issuenum{1}
\articlenumber{0}
\pubyear{2022}
\copyrightyear{2022}
 \externaleditor{Academic Editor:{ Yi-Zhong Fan } 
} \datereceived{29 July 2022 } 
\dateaccepted{25 August 2022} 
\datepublished{} 

\hreflink{https://doi.org/} 

\usepackage{longtable}
\usepackage{makecell}
\usepackage{aas_macros}

\Title{TeV Dark Matter Searches in the Extragalactic Gamma-Ray Sky}

\TitleCitation{TeV Dark Matter Searches in the Extragalactic Gamma-Ray Sky}

\Author{Moritz Hütten $^{1,}$*\orcidA{} and Daniel Kerszberg $^{2,}$*\orcidB{}}

\AuthorNames{Moritz Hütten and Daniel Kerszberg}

\AuthorCitation{Hütten, M.; Kerszberg, D.}

\address{%
$^{1}$ \quad Institute for Cosmic Ray Research, The University of Tokyo,  Kashiwa, Chiba 277-8583, Japan\\
$^{2}$ \quad Institut de F\'{i}sica d'Altes Energies (IFAE), The Barcelona Institute of Science and Technology (BIST),\linebreak E-08193 Bellaterra, Barcelona, Spain \\
}
\corres{Correspondence: huetten@icrr.u-tokyo.ac.jp (M.H.); dkerszberg@ifae.es (D.K.)}

\abstract{High-energetic gamma rays from astrophysical targets constitute a unique probe for annihilation or decay of heavy particle dark matter (DM). After several decades, diverse null detections have resulted in strong constraints for DM particle masses up to the TeV scale. While the gamma-ray signature is expected to be universal from various targets, uncertainties of astrophysical origin strongly affect and weaken the limits. At the same time, spurious signals may originate from non-DM related processes. The many gamma-ray targets in the extragalactic sky being searched for DM play a crucial role to keep these uncertainties under control and to ultimately achieve an unambiguous DM detection. Lately, a large progress has been made in combined analyses of TeV DM candidates towards different targets by using data from various instruments and over a wide range of gamma-ray energies. These approaches not only resulted in an optimal exploitation of existing data and an improved sensitivity, but also helped to level out target- and instrument-related uncertainties. This review gathers all searches in the extragalactic sky performed so far with the space-borne \textit{Fermi}-Large Area Telescope, the ground-based imaging atmospheric Cherenkov telescopes, and the High-Altitude Water Cherenkov Gamma-Ray Observatory (HAWC). We discuss the different target classes and provide a complete list of all analyses so far.
}


\newcommand{\hess}{{\small H.E.S.S.}}

\newcommand{\beq}{\begin{equation}}
\newcommand{\eeq}{\end{equation}}
\newcommand{\balign}{\begin{align}}
\newcommand{\ealign}{\end{align}}

\newcommand{\lcdm}{{\ifmmode \Lambda{\rm CDM} \else $\Lambda{\rm CDM}$\fi}}

\newcommand{\Msol}{\ensuremath{\rm M_\odot}}

\newcommand{\sigmav}{\ensuremath{\langle \sigma v\rangle}}

\newcommand{\dd}{\ensuremath{\mathrm{d}}}
\newcommand{\fermi}{{\em Fermi}-LAT}

\begin{document}

\section{Introduction\label{sec:intro}}

For about a hundred years, powerful evidence has been accumulating that about 80\% of the cosmic mass budget is yet of unknown nature, coined the dark matter (DM)~\cite{2012AnP...524..479B}. Among the many proposed physical origins of DM, a promising hypothesis is the existence of weakly interacting massive particles (WIMPs) on the scale of GeV to TeV within some theory beyond the current Standard Model of particle physics~\cite{2018RPPh...81f6201R}. However, despite the tremendous experimental efforts to test the WIMP hypothesis and various other DM candidates, the nature of DM is still elusive~\cite{10.1093/ptep/ptac097}. 

Searches in astrophysical gamma rays play a crucial role in these efforts. The use of astrophysical gamma rays to probe DM candidates was already proposed in the late 1970s in the GeV regime: possible signatures from annihilation of massive particles were first discussed for the extragalactic diffuse gamma-ray background~\cite{1978ApJ...223.1015G} and the center of the Milky Way~\cite{1978ApJ...223.1032S} to be searched in the data of early satellite gamma-ray detectors. Searches were followed up with the EGRET satellite~\cite{2004NuPhS.134..103B} and lately perfected with unprecedented sensitivity by the \textit{Fermi-}Large Area satellite Telescope (\fermi{})~\cite{2016PhR...636....1C}.

\begin{figure}[t] 
\begin{adjustwidth}{-\extralength}{0cm}
\centering
\includegraphics[width=\linewidth]{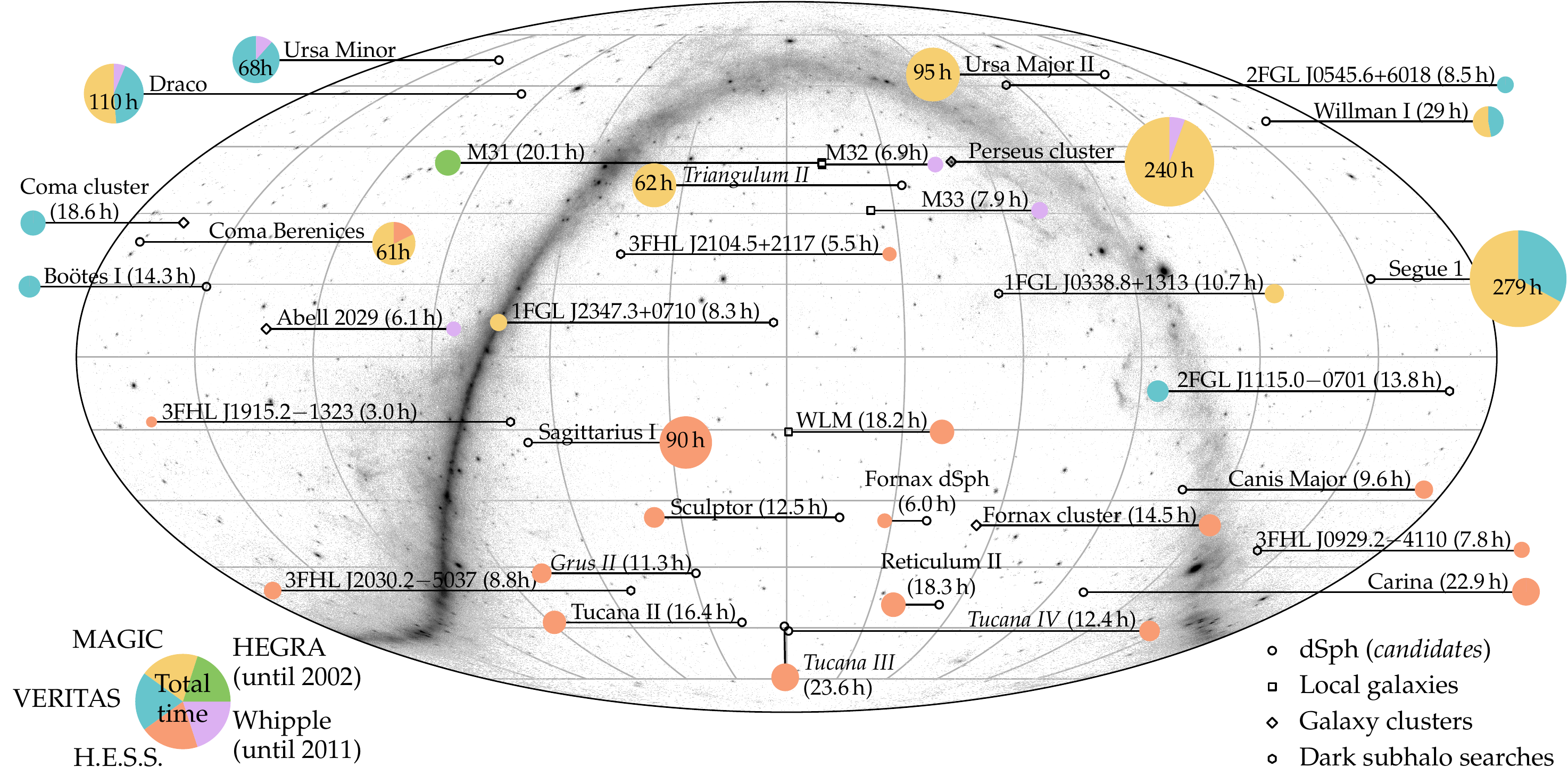}
\end{adjustwidth}
\caption{All published exposures so far of IACTs on DM targets in  the extragalactic sky (in RA-DEC coordinates). The pie chart sizes illustrate the total observation time spent on each target. In total, the map shows 1329 h  of observation spent on 32 targets, out of which $716\,$h are taken by MAGIC, $292\,$h by H.E.S.S, $271\,$h by VERITAS, $50\,$h by Whipple, and $20\,$h by HEGRA. The background shows for illustration the \fermi{} counts map after 705 weeks (13.5 years) above 1 GeV. Note that data on galaxies and galaxy clusters were taken for multi-science purposes, while dSph targets were observed exclusively in hunt for DM. Targets written in italics are still unconfirmed dSph candidates.}
\label{fig:skymap_iact}
\end{figure}

\begin{figure}[t] 
\begin{adjustwidth}{-\extralength}{0cm}
\includegraphics[width=\linewidth]{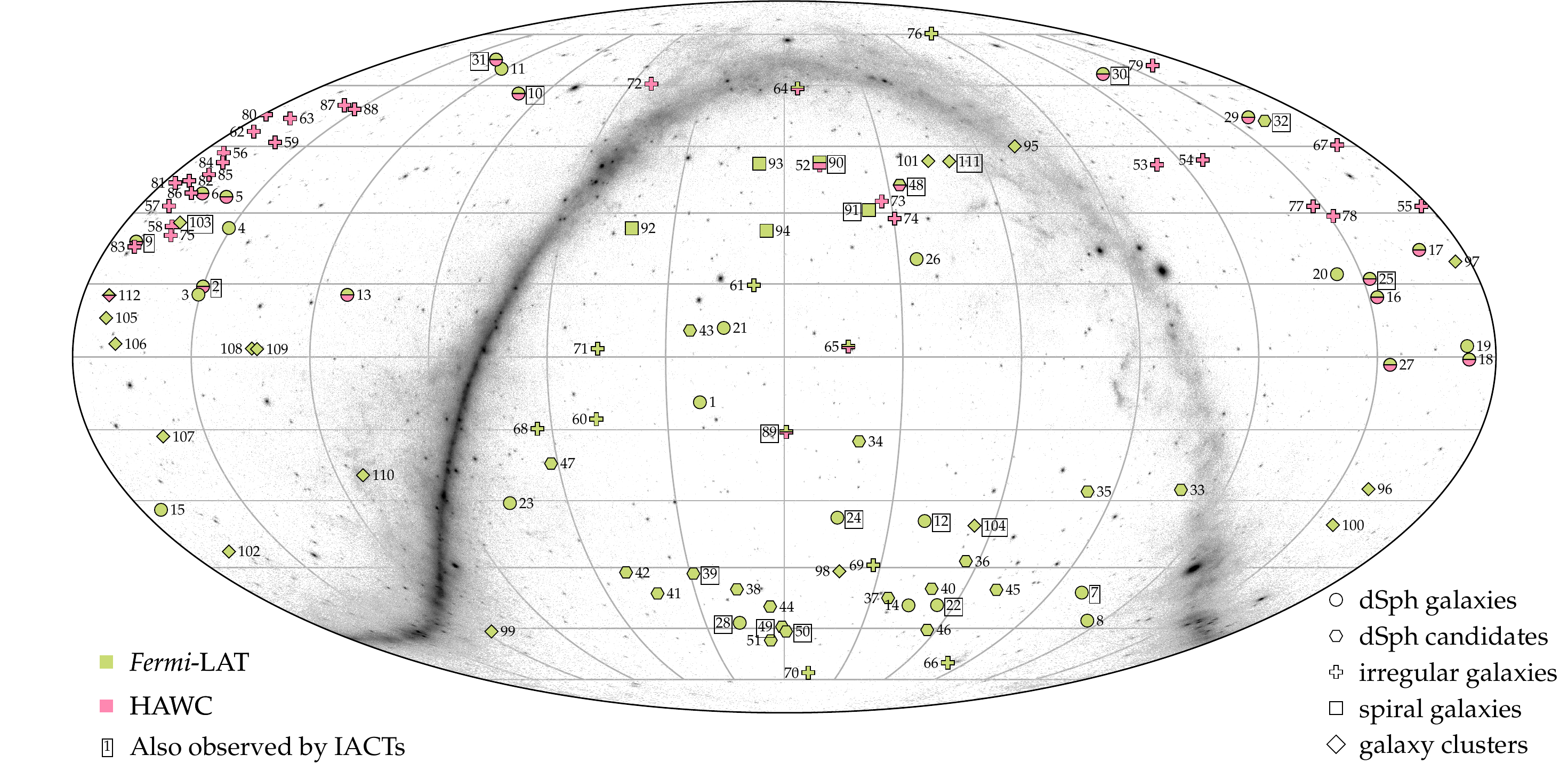}
\end{adjustwidth}
\caption{Searches for WIMP DM done on targets in the extragalactic sky with the data by \fermi{} (green, top half of divided markers) and HAWC (magenta, bottom markers) detectors. The dozens of dark subhalo candidates among the \fermi{} and HAWC unidentified sources are not included in this picture. Only the 18 ``most popular'' galaxy clusters are shown, i.e., individually analyzed for DM in at least two peer-reviewed publications in our literature search. The numbers next to the indicators of the targets' positions refer to the source names listed in Table~\ref{tab:fermihawc-targetsref}. Numbers within a box indicate targets also observed with IACTs as shown in Figure \ref{fig:skymap_iact}.}
\label{fig:skymap_fermihawc}
\end{figure}

Within supersymmetric extensions of the Standard Model, new particles on the TeV scale are favored as DM candidates~\cite{1982PhRvL..48..223P}, and gamma-ray searches close to the spectral endpoint at TeV energies can naturally provide an important puzzle piece for the identification of DM. At these energies, DM signals might be better distinguishable from decreased astrophysical backgrounds and may show prominent spectral features. However, such energies can hardly be probed by satellite detectors. Searching for DM particles in the TeV regime with ground-based Atmospheric Cherenkov telescopes was first discussed by~\cite{1992PhLB..293..149U} in 1992, the same year as the first extragalactic TeV gamma-ray source was detected by the Whipple telescope~\cite{1992Natur.358..477P}. First pioneering searches for DM annihilation with the Whipple 10-meter telescope and the HEGRA array were conducted a decade later towards nearby galaxies~\cite{2008ApJ...678..594W,2003A&A...400..153A} and galaxy clusters~\cite{2006ApJ...644..148P}. DM searches with the current generation of imaging atmospheric Cherenkov telescopes (IACTs) were already detailed in the late 1990s for probing the Galactic center (GC) region~\cite{1998APh.....9..137B} and external galaxies~\cite{1999PhRvD..61b3514B}. While the GC region promises a bright gamma-ray signal from DM annihilation, it suffers from background emission up to TeV energies~\cite{2004ApJ...608L..97K}, and major uncertainties about the expected DM signal were discussed~\cite{2004PhRvD..69l3501E}. These caveats resulted in an increased interest in alternative targets. To date, around 2500~h have been invested by IACTs in the search for DM on the TeV scale, with more than half of this time distributed on dozens of targets in the extragalactic sky.\endnote{In this review, we consider all targets outside the Galactic plane. However, we omit DM searches in high-latitude globular clusters. A recent review including DM searches in the inner Galaxy, Galactic globular clusters, and intermediate black holes, as well as their particular challenges, can be found in~\cite{2021arXiv211101198D}.} In Figure \ref{fig:skymap_iact}, we illustrate the observations on all these targets outside the Galactic plane with the various IACTs, along with the respective exposures. Besides IACTs, also water Cherenkov detectors like the High-Altitude Water Cherenkov Observatory (HAWC), similar to \fermi{} constantly surveying a large fraction of the sky, are excellent instruments for TeV DM searches. In Figure \ref{fig:skymap_fermihawc}, we show more than 100 DM targets in the extragalactic sky analyzed with HAWC and in the public data of \fermi{}.

This contribution reviews the current status of all these DM searches in the extragalactic sky, what we have learnt from them, and their particular challenges. In Section~\ref{sec:indirect_foundations}, we introduce the basic concepts of indirect DM searches. In Sections~\ref{sec:dwarfs}--\ref{sec:clusters}, we present the results of TeV DM searches obtained so far on the various targets. In Section~\ref{sec:mwl}, we draw a focus on multi-target, multi-instrument, and multi-messenger searches. Additionally, not only WIMP DM may show gamma-ray signatures---also axion-like particles (ALPs) and primordial black holes (PBHs) are viable DM candidates that can be searched for with gamma rays. While ALPs are covered in the contribution~\cite{2022Galax..10...39B} to this Special Issue, we will discuss searches for PBHs in Section~\ref{sec:pbhs}. Finally, we give a short outlook and conclude in Section~\ref{sec:conclusions}.

\section{Gamma-Ray Signals from WIMP Dark Matter\label{sec:indirect_foundations}}

WIMP DM may be detected in gamma rays following annihilations or decays of relic DM particles in today's Universe~\cite{2000RPPh...63..793B}. DM particles may annihilate or decay into the lighter particles contained in the known Standard Model of Particle Physics, eventually generating electromagnetically interacting massive particles and high-energetic photons. While such processes could in principle be probed in current laboratory detectors~\cite{2022PTEP.2022a3F01M}, stronger signals are expected for telescopes observing astrophysical targets~\cite{2015PNAS..11212264F}. This strategy to probe the nature of DM in astrophysical objects is called indirect search
for DM, in contrast to direct
searches for DM in laboratories~\cite{2022RPPh...85e6201B} and collider experiments~\cite{doi:10.1146/annurev-nucl-101917-021008}. In case of astrophysical gamma-ray signals, most radiation is assumed to be produced directly at the place of annihilation\endnote{Gamma rays can additionally be produced in the astrophysical surrounding by secondary processes like inverse Compton scattering of high-energetic $e^+/e^-$ produced after DM annihilation or decay~\cite{2009NuPhB.821..399C,2009JCAP...07..020P}.} and the electromagnetically neutral photons travel straight (in a curved spacetime) to the observer. This allows a direct mapping of the projected cosmic DM density distribution. From numerical simulations, today's DM distribution is expected to show a high density contrast, with self-similar structures over many mass scales~\cite{2019Galax...7...81Z}. In particular, these structures include the DM halos surrounding galaxies and galaxy clusters that provide the gravitational evidence for DM. This facilitates observations by IACTs pointed towards very massive or close-by locations of such DM overdensities. For a DM density distribution $\rho$ (given in co-moving spatial coordinates) at redshift $z$ in a region $\Delta z/z\ll 1$, the gamma-ray flux from  annihilation can be written as:
\begin{align}
\small
\label{eq:flux_1halo_ann} 
\frac{\Phi_{\rm ann}}{\dd E_{\gamma}}(E_{\gamma},\Delta\Omega)
= \frac{\sigmav}{8\pi\,m_{\mathrm{DM}}^2} \, \left.\frac{\dd N}{\dd E}\right|_{E=(1+z)E_{\gamma}} \times e^{-\tau(z,\, E_{\gamma})}
\times \underbrace{(1+z)^3\, \int\displaylimits_0^{\Delta \Omega} \int\displaylimits_{\rm l.o.s.} \,\rho(l,\Omega)^2\,\dd l \,\dd \Omega}_{=:\, J_{\rm ann}}
\,.
\end{align}
{Here,} 
 it is assumed that the DM is made of Majorana particles, so that DM particles are identical to their antiparticles. The signal from decaying DM reads as:
\begin{align}
\small
\label{eq:flux_1halo_dec} 
\frac{\Phi_{\rm decay}}{\dd E_{\gamma}}(E_{\gamma},\Delta\Omega)
=\frac{1}{4\pi\,t_\mathrm{DM}\,m_{\mathrm{DM}}} \, \left.\frac{\dd N}{\dd E}\right|_{E=(1+z)E_{\gamma}} 
\times e^{-\tau(z,\, E_{\gamma})} 
\times \underbrace{\int\displaylimits_0^{\Delta \Omega} \int\displaylimits_{\rm l.o.s.} \,\rho(l,\Omega)\,\dd l \,\dd \Omega}_{=:\, J_{\rm dec}}\,.
\end{align}

In these equations, $m_{\mathrm{DM}}$ denotes the mass of the WIMP, $E_\gamma$ the observed gamma-ray energies, $\dd N/ \dd E$ the number of photons per energy resulting from a single annihilation or decay process, and l.o.s the integration along the line of sight, the co-moving radial coordinate $l$. $\Delta \Omega$ is the viewing cone. \sigmav{} and $t_{\text{DM}}$ denote the velocity-averaged annihilation cross section and WIMP lifetime, respectively. For far-away gamma-ray sources, the flux is attenuated by absorption of photons on the extragalactic background light with the gamma-ray optical depth $\tau$. While this attenuation is negligible for sources in the Local volume ($z\lesssim 10^{-3}$) and sub-TeV energies, it may play a significant role for sources at higher redshift and/or if the WIMP mass is located at higher energies. In particular, for DM models on the PeV scale beyond the unitary bound~\cite{2019PhRvD.100d3029S}, attenuation on the cosmic microwave background (CMB) can significantly suppress the signal on subgalactic scales~\cite{2018JCAP...04..060B}.

\begin{figure}[H]
\includegraphics[width=0.73\linewidth]{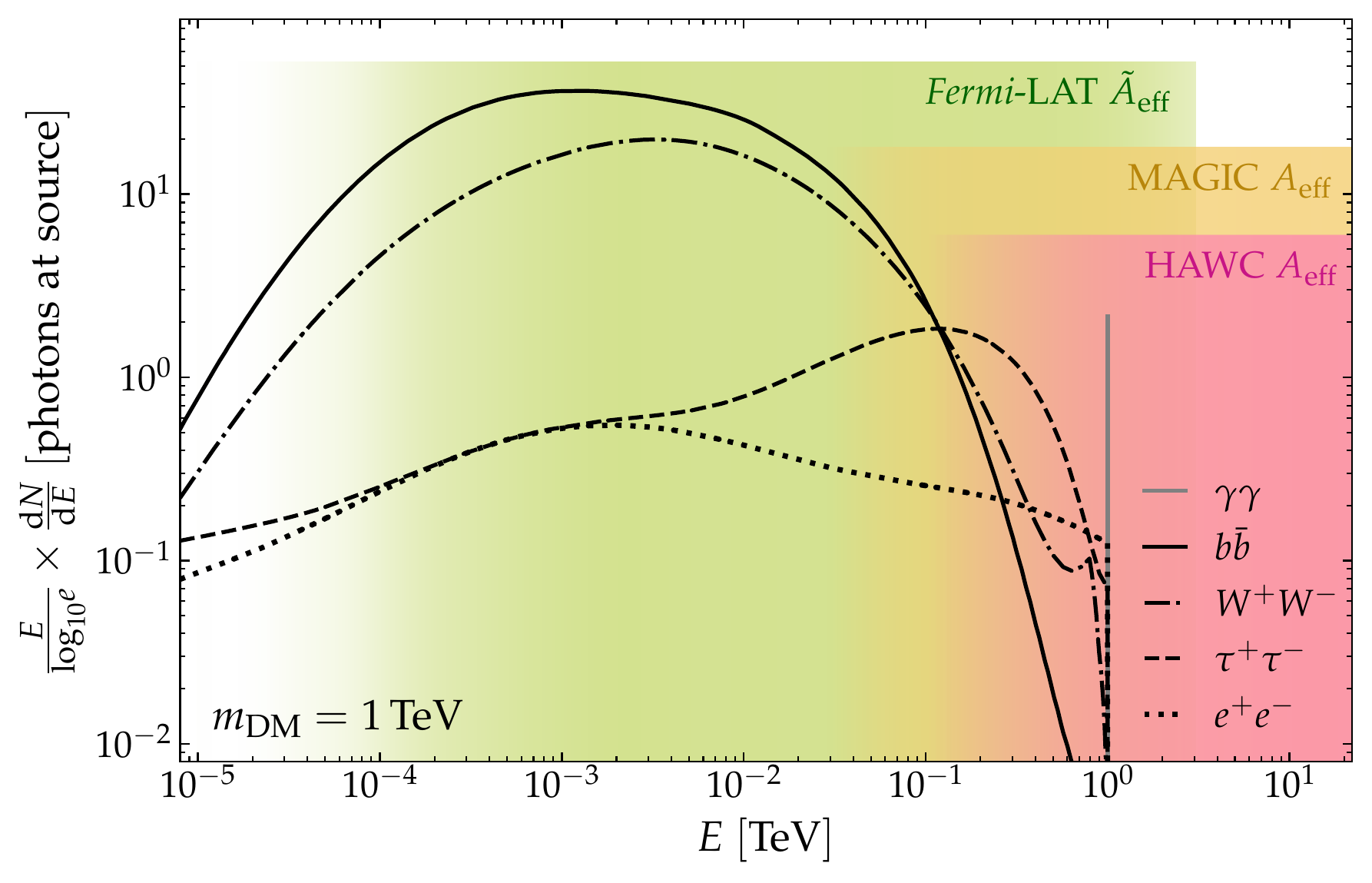}
\includegraphics[width=0.73\linewidth]{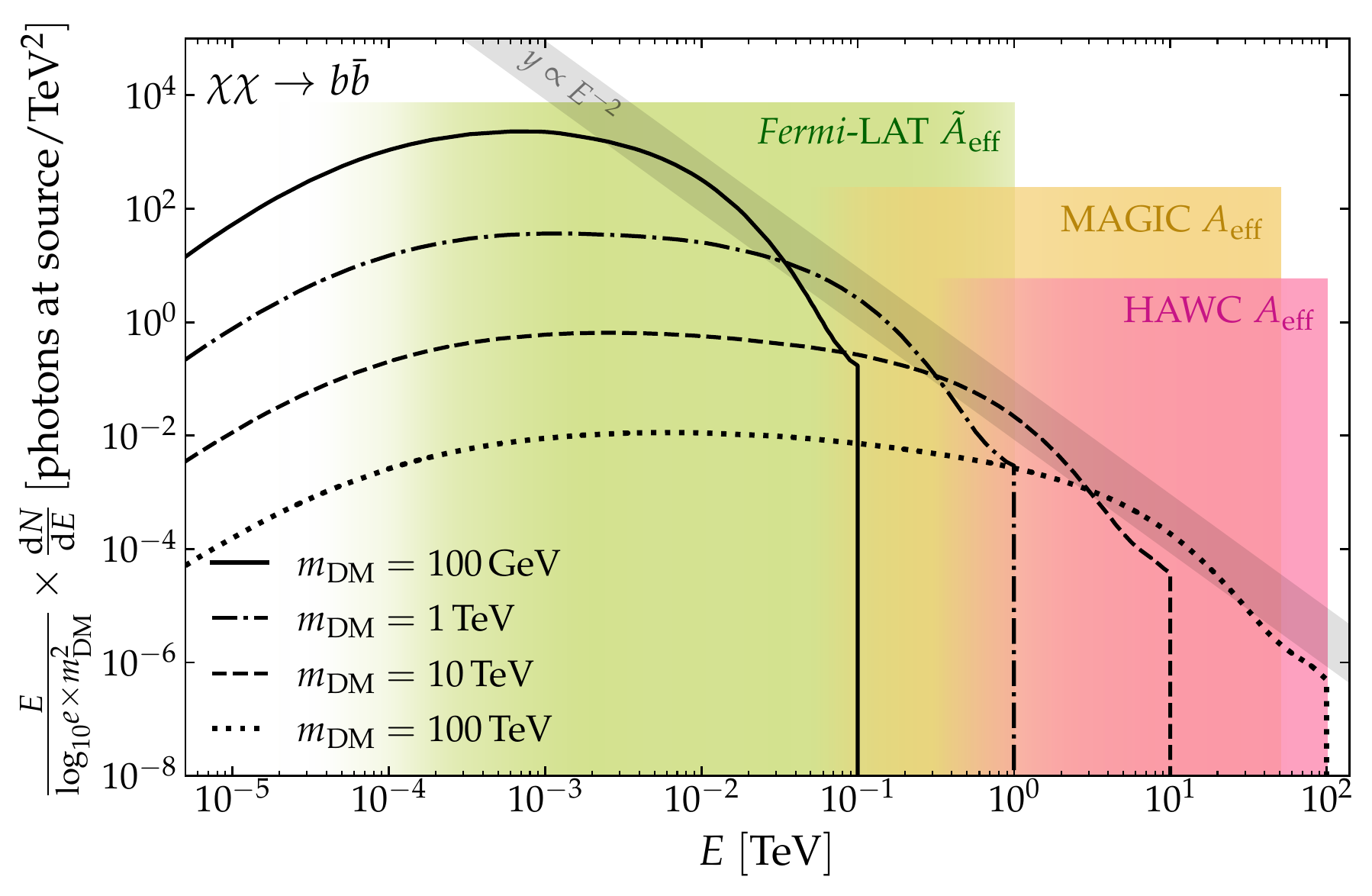}
\caption{\textit{Top~panel:} Gamma-ray spectra for different annihilation channels and a fixed DM mass. The vertical axis is scaled such that the number of photons in an energy interval can be read by eye when integrating over one order of magnitude (e.g., for annihilations into $b\bar{b}$ quarks, approximately \mbox{35 photons} are created per annihilation between 1 and 10 GeV). \textit{Bottom panel:} Different DM particle masses for a given annihilation channel. The vertical axis is scaled by $m_{\mathrm{DM}}^{-2}$, indicating that about \mbox{100 times} less photons reach the observer when the DM  mass increases by a factor 10. The sensitivity range of \fermi{}~\cite{2018arXiv181011394B} (P8R3\_SOURCE\_V3 all events), MAGIC~\cite{2016APh....72...76A} (0--30 $^\circ$ after cuts), and HAWC~\cite{2017ApJ...843...39A} (with $\gamma$/h sep.) is illustrated by their effective areas (i.e., solid-angle-averaged acceptance for \fermi{}).}
\label{fig:dNdE_spectra}
\end{figure}

\begin{figure}[H]
\includegraphics[width=\linewidth]{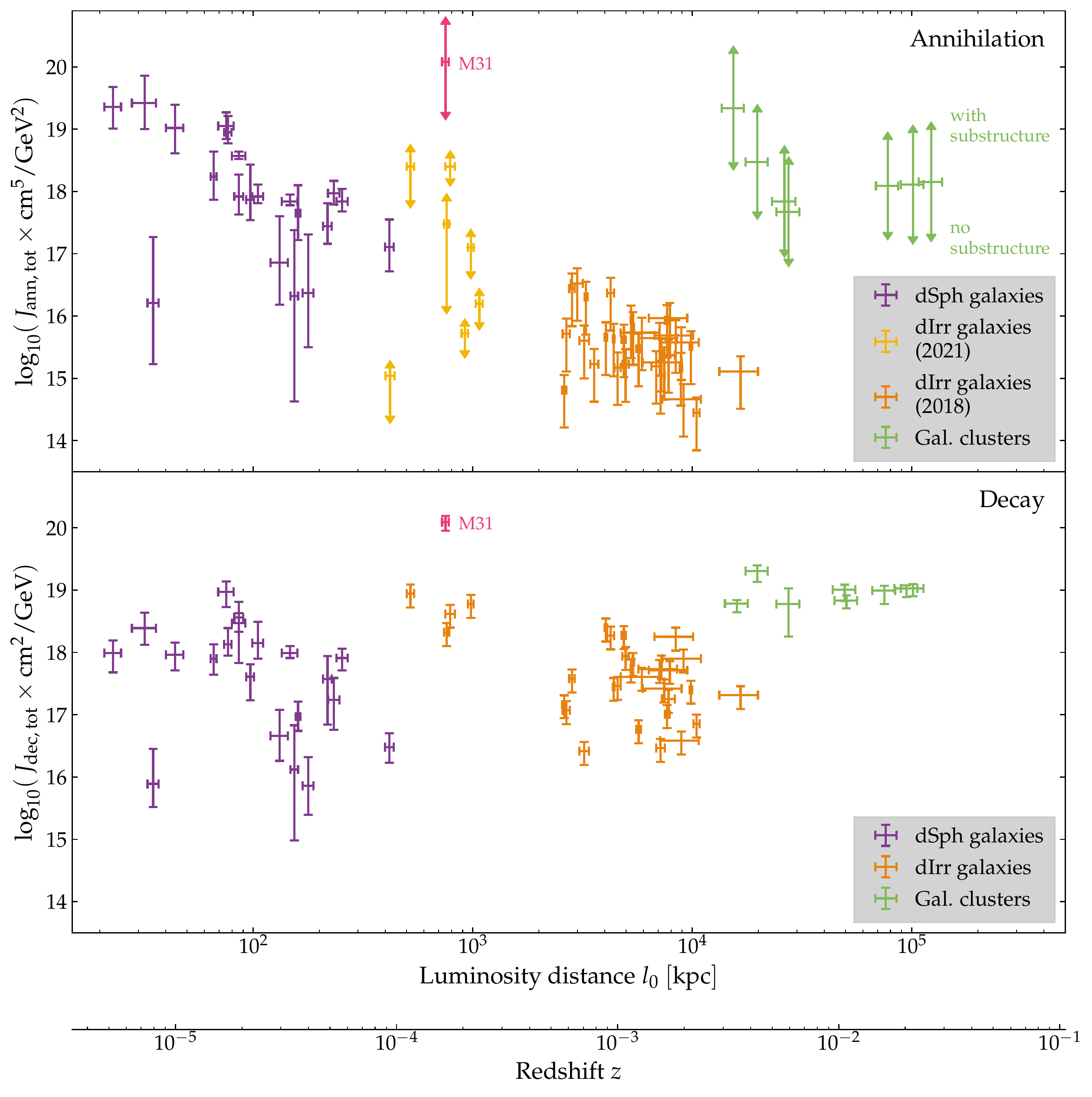}
\caption{\textit{Top~panel:} $J$-factors according to Equation~(\ref{eq:flux_1halo_ann}) for DM annihilation as a function of distance $l_0$ to the objects' centers. dSph galaxies (purple): values from~\cite{2015ApJ...801...74G}. dIrr galaxies: values from~\cite{2021PhRvD.104h3026G} (light orange, 2021) and~\cite{2018PhRvD..98h3008G} (dark orange, 2018), distance values additionally from~\cite{1991AJ....102..130V,2012AJ....144....4M}. M31 (magenta):~\cite{2018JCAP...06..043A}. Galaxy clusters (green): from~\cite{2011JCAP...12..011S}, with a $12\%$ distance uncertainty according to~\cite{1997MNRAS.289..847D}. Where error bar caps are shown as arrows, the uncertainty is assumed to be dominated by the uncertain signal boost from DM substructure. \textit{Bottom panel:} same for decay according to Equation~(\ref{eq:flux_1halo_dec}). dSph galaxies (purple): values from~\cite{2015ApJ...801...74G}. dIrr galaxies (orange): values from~\cite{2019ICRC...36..520H}, with a $40\%$ error according to~\cite{2018PhRvD..98h3008G}. M31 (magenta):~\cite{2018JCAP...06..043A}. Galaxy clusters (green): from~\cite{2012JCAP...01..042H}.}
\label{fig:jfactors}
\end{figure}

The integrals over the density distribution in Equations~(\ref{eq:flux_1halo_ann}) and (\ref{eq:flux_1halo_dec}) are called the \textit{astrophysical factor}, usually denoted in short $J$-factor (or sometimes $D$-factor in case of decay). In case of annihilation, the signal scales with the number of DM particles times the DM density, as two particles from the same density distribution are involved in the annihilation process. For decay in turn, the signal scales merely with the number of particles in the viewing cone $\Delta \Omega$. Further, for annihilation, the flux scales with $m_{\mathrm{DM}}^{\;-2}$ (see Figure \ref{fig:dNdE_spectra}, bottom panel), while only with $m_{\mathrm{DM}}^{\;-1}$ for decaying DM. We will later see that this different scaling for decay allows for strong gamma-ray constraints on $t_{\text{DM}}$ for $m_{\mathrm{DM}}$ larger than several TeV. The $J$-factor usually constitutes the largest uncertainty in indirect DM searches. This holds especially in case of annihilation, for which it depends on the density subclustering of a DM halo~\cite{2019Galax...7...68A}. For the same reason, the factor $(1+z)^3$ in front of the density integral in Equation~(\ref{eq:flux_1halo_ann}) expresses that a higher density in proper coordinates at the time when the annihilations took place boosts the annihilation rate.\endnote{This factor is usually neglected due to the large uncertainties in the density distribution entering the $J$-factor.} For most sources, their angular extension $\dd J/\dd \Omega$ is expected to be strongly peaked towards the DM halo center and appears almost point-like to current-generation gamma-ray telescopes. The extension is larger in the case of decay, and being able to identify the extended emission profile may help to pinpoint the DM origin of a signal. In Figure \ref{fig:jfactors}, we show the total $J$-factors, $J_{\mathrm{tot}}= J(\Delta\Omega\rightarrow 4\pi)$,\endnote{The size of a DM halo is  usually limited by a defined virial or characteristic overdensity radius. Accordingly, the integral of Equations~(\ref{eq:flux_1halo_ann}) and (\ref{eq:flux_1halo_dec}) vanishes at angular distances from the halo center larger than this radius.} for both annihilation and decay, and for different source classes. It can be seen the larger uncertainties of $J_{\rm ann}$ compared to $J_{\rm dec}$, especially when targets are expected to host a significant number of DM substructure (error bar caps marked by arrows). Note that comparing the merit of different targets is more complex, as some targets are affected by background gamma-ray emission or are very extended; the latter case poses a challenge in particular for instruments with a small field of view and residual isotropic backgrounds.

Additionally, the expected spectral energy distribution of gamma rays $\dd N/ \dd E$ is crucial for the DM interpretation of a signal. At the astrophysical source, it depends on the composition of particles produced after DM annihilation or decay. This branching composition, \mbox{$\dd N/ \dd E = \sum_i f_i (\dd N/ \dd E)_i$} into different channels $i$, is largely unknown and depends on the specific DM particle model. However, given a specific channel of annihilations or decays into Standard Model particles, the spectra can be robustly computed and are shown in \mbox{Figure \ref{fig:dNdE_spectra}~(top).} For interpreting null measurements, indirect DM searches usually scan individual annihilation or decay channels and derive limits under the assumption $f_i =1$ for each channel. The actual limits may indeed be roughly bracketed between the limit predicted by a soft gamma-ray spectrum (i.e., few gamma rays arriving in the instrument's energy range) and that by a hard channel. Assuming the DM to be cold, i.e., at rest at annihilation, all gamma-ray spectra share a sharp cut-off at the DM particle mass; momentum conservation guarantees that no gamma ray can be produced being more energetic than the rest mass of a single DM particle. In case of decay, this cut-off occurs at half the DM particle mass. After DM annihilation or decay into Standard-model matter, most of the high-energetic gamma-ray emission is produced by $\pi_0$ decays in hadronic cascades. Therefore, below the cut-off, most channels result in a rather featureless continuum spectrum (see Figure \ref{fig:dNdE_spectra},~top panel). Distinguished from these continuum spectra are cases where one or several gamma rays are directly produced in the annihilation or decay and carry away a large fraction of the DM particles' energy. This includes direct annihilation or decay into a gamma-ray pair, a gamma ray and another vector boson, or three-body processes including gamma rays like virtual internal bremsstrahlung. Such emission would not only provide compelling evidence for some exotic new physics, but also make it easier to detect for background-dominated detectors. However, such processes are for most DM models suppressed by orders of magnitude compared to annihilation or decay into pairs of heavy leptons or gauge bosons~\cite{1997NuPhB.504...27B,2007PhLB..646...34H,2020JHEP...03..030B}.

\newpage
\section{Dwarf Spheroidal Galaxies\label{sec:dwarfs}}

Dwarf spheroidal galaxies (dSphs) are low-luminosity satellite galaxies of the Milky Way or Andromeda (at least those detected to this date) and form the majority of the Local group galaxies~\cite{2013AJ....145..101K}. Their name is due to their relatively small size, containing typically not more than $10^3$--$10^7$ stars~\cite{2018RPPh...81e6901S}, and to the fact that these stars follow more or less a spherical distribution. They are formed in the DM subhalos surrounding galaxies with enough baryonic matter to start and sustain stellar formation. With DM-dominated masses of about $10^6$--$10^{9}\,\Msol$~\cite{2015ApJ...801...74G,2015MNRAS.453..849B}, they are the DM targets with the largest mass-to-light (M/L) ratio detected to this date, although they could be surpassed by the predicted but yet still undetected dark DM subhalos (see Section~\ref{sec:subhalos}). dSphs are old objects that only had a few generations of stars. As little to no star formation is occurring in these systems today, dSphs are considered as very clean targets for indirect DM searches and are expected to host no background gamma-ray emitters. dSphs typically orbit the Milky Way or the Andromeda galaxy at a few dozen to a few hundred kiloparsecs. In the Milky Way, this closely located to Earth, it allows to resolve the spectra of their stars individually and to estimate their DM profiles with a relative good precision in most of the cases, allowing relatively robust predictions of their $J$-factors. A relatively low amount of DM substructure is expected in dSphs, translating in an negligible boost of gamma-ray emission from DM annihilation~\cite{2019Galax...7...68A}.

\begin{figure}[H]
\includegraphics[width=\linewidth]{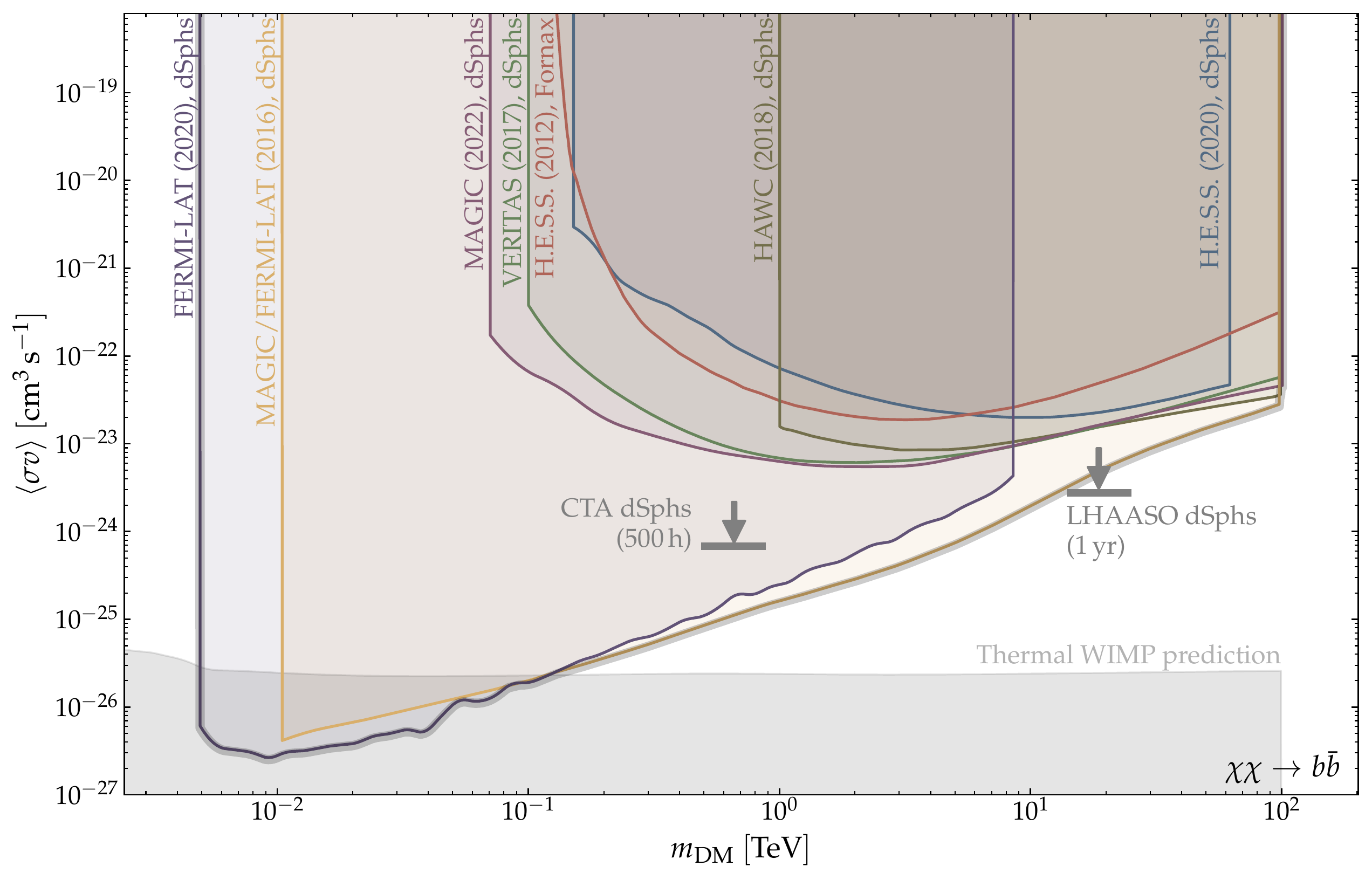}
\caption{Best limits on the cross section of WIMP annihilation into $b\bar{b}$ quarks from targets in the extragalactic sky. From left to right: constraints from the analysis of the data towards 27 stacked dSphs with \fermi{}~\cite{2020JCAP...02..012H} (freq. glob. fit with 1 DOF), combined analysis of 15 \fermi{} stacked dSphs and MAGIC Segue 1 observations~\cite{2016JCAP...02..039M}, analysis of 4 stacked dSphs observed with MAGIC~\cite{2022PDU....3500912A} and 5 stacked dSphs with VERITAS~\cite{2017PhRvD..95h2001A}, of the Fornax galaxy cluster with \hess{}~\cite{2012ApJ...750..123A} (NFW RB02, high sub., $\theta = 1^\circ$), analysis of 15 stacked dSphs with HAWC~\cite{2018ApJ...853..154A}, and 5 stacked dSphs observed with \hess{}~\cite{2020PhRvD.102f2001A}. Gray arrows indicate the projected sensitivity of future instruments at their best sensitivity in the $m_{\rm DM}$ parameter space for 500$\,$h observations of Segue I dSph with CTA~\cite{2019scta.book.....C} and 1~year combined dSph observations with LHAASO~\cite{2019PhRvD.100h3003H}.
Analyses assumed different $J$-factors for the same targets. Thermal WIMP cross section prediction: analytic calculation from~\cite{Huetten2017Prospects}.}
\label{fig:bestlimits_ann}
\end{figure}

The observed number of dSphs and their DM profile is not without posing questions though. The same simulations that predict the existence of such DM-dominated objects orbiting the Milky Way also predict cuspy profiles of the DM density profile near the centers~\cite{2008MNRAS.391.1685S,2008Natur.454..735D}. In turn, spectroscopic measurements of dSphs show a variety of DM profiles from cores to cusps, a discrepancy known as the ``core-cusp'' problem~\cite{2018MNRAS.474.1398G}. Furthermore, simulations suggest the existence of more and heavier satellites than what has been observed so far, a problem known as the ``missing satellites'' problem~\cite{1999ApJ...522...82K,2007ApJ...669..676S}. While the problem is claimed to be largely understood~\cite{2007ApJ...670..313S,2018PhRvL.121u1302K,2019MNRAS.487.5799R}, one explanation is that some dSphs lack baryonic mass (or lost some via unaccounted tidal effects) which could explain why their luminosity is too low to have yet observed them. However, this is in contradiction to what is expected for such massive objects which should host visible stars. This contradiction is known as the ``too big to fail'' problem~\cite{2019MNRAS.490..231K}: these objects should be  too big to fail to host star formation, at least in our current understanding of galaxy evolution and star formation.

dSphs were extensively used for the search of DM and adopted as dedicated targets for IACTs. Instruments with a large field of view such as \fermi{} and HAWC observed them as part of their continuous sky monitoring. The IACTs used their sensitivity to accumulate deep exposures on the promising targets shown in Figure \ref{fig:skymap_iact}: Draco and Ursa Minor by Whipple in 2008~\cite{2008ApJ...678..594W}, Sagittarius~I in 2008 by H.E.S.S.~\cite{2008APh....29...55A}, Sculptor and Carina in 2011 by H.E.S.S.~\cite{2011APh....34..608H}, Canis Major (still classified as a dSph candidate to this day due to its disputed nature, see e.g.~\cite{2021MNRAS.501.1690C}) in 2009 by H.E.S.S.~\cite{2009ApJ...691..175A}, Draco in 2008 by MAGIC~\cite{2008ApJ...679..428A}, Willman 1 in 2009 by MAGIC~\cite{2009ApJ...697.1299A}, Segue 1 in 2011 by MAGIC~\cite{2011JCAP...06..035A}, Segue 1 in 2014 with stereoscopic observations by MAGIC~\cite{2014JCAP...02..008A}, Ursa Major II in 2018 by MAGIC~\cite{2018JCAP...03..009A}, Triangulum II in 2020 by MAGIC~\cite{2020PDU....2800529A}, Willman 1, Draco, Ursa Minor, and Boötes I by VERITAS in 2010~\cite{2010ApJ...720.1174A}, and a deep exposure with VERITAS on Segue 1 in 2012~\cite{2012PhRvD..85f2001A}. At first, IACTs concentrated on these single targets and on deriving limits for single promising objects. At the same time, the lower sensitivity of \fermi{} and HAWC, with however steadily growing exposure and without the need of choosing a target to point at, let them focus on multi-target analyses from the beginning: the \fermi{} collaboration has published constraints on DM annihilation based on 11~months of data on 14~dSphs in 2010~\cite{2010ApJ...712..147A}, 2~years of 10~dSphs in 2011~\cite{2011PhRvL.107x1302A}, 4~years of 25~dSphs in 2014~\cite{2014PhRvD..89d2001A}, 6~years of 25~dSphs in 2015~\cite{2015PhRvL.115w1301A}, and 6~years of 45~objects in 2017~\cite{2017ApJ...834..110A}. The latest limits by~\cite{2020JCAP...02..012H} using 11 years of the public \fermi{} data are shown in Figure \ref{fig:bestlimits_ann} for annihilations into $b\bar{b}$ quarks. It can be seen that constraints are done with \fermi{} up to DM masses of 10~TeV. In 2018, HAWC published an analysis of 15~dSphs with 507 days of data ~\cite{2018ApJ...853..154A}. The stacking approach pursued by \fermi{} and HAWC has the advantage of maximizing the sensitivity of searching a universal DM signal and was gradually also adopted by IACTs: H.E.S.S. published analyses of $\sim$140~h on 5~dSphs (Carina, Coma Berenices, Fornax, Sagittarius~I, Sculptor) in 2014~\cite{2014PhRvD..90k2012A} (continuum) and 2018~\cite{2018JCAP...11..037A} (line emission) and of $\sim$$82$~h on  Reticulum II, Tucana II, Tucana III, Tucana IV and Grus II in 2020~\cite{2020PhRvD.102f2001A}. VERITAS published $\sim$230~h on 5~dSphs (Boötes I, Draco, Segue 1, Ursa Minor, Willman 1) in 2017~\cite{2017PhRvD..95h2001A}, and MAGIC $\sim$354~h on 4~dSphs (Coma Berenices, Draco, Segue 1, Ursa Major II) in 2022~\cite{2022PDU....3500912A}. In 2016, \fermi{} and MAGIC published a joint analysis of 15 dSphs observed by \fermi{} plus the $\sim$160~h accumulated by MAGIC on Segue~1~\cite{2016JCAP...02..039M}. That analysis  inaugurated multi-instrument gamma-ray searches for DM annihilation in dSphs (see Section~\ref{sec:mwl}). In Figure \ref{fig:bestlimits_ann}, we include all the latest combined searches by IACTs, compared against the \fermi{} and HAWC results. It can be seen that, while \fermi{} provides the strongest constraints in the GeV regime, competitive limits are obtained from IACTs and HAWC for DM masses above tens of TeV. In particular, the multi-instrument combination~\cite{2016JCAP...02..039M} still provides one of the strongest constraints for DM masses on the TeV scale.\endnote{Note, however, that these compared analyses do not use consistent $J$-factor values for the same targets.} While dSphs were so far mostly studied to search for DM annihilation, \cite{2016PhRvD..93j3009B} have analyzed 6~years of \fermi{} towards 19~dSphs, providing strong constraints on the lifetime of GeV DM (Figure \ref{fig:bestlimits_dec}). This illustrates the potential of dSphs also for searching for decaying DM.
\vspace{-6pt}

A complete list of the analyzed dSphs searched so far for DM annihilation or decay with \fermi{} and HAWC is given in Table~\ref{tab:fermihawc-targetsref}, as completion of a similar table gathering all IACT observations in~\cite{2021arXiv211101198D}. Furthermore, another recent review describing in depth the statistical analysis used in the search of DM in the direction of dSphs with \fermi{}, HAWC, H.E.S.S., MAGIC, and VERITAS, can be found in~\cite{2020Galax...8...25R}. Note that while a worsening of most of the dSph constraints by a factor of 2 to 7 had been recently claimed~\cite{2020PhRvD.102f1302A}, a most recent re-evaluation suggests the validity of earlier $J$-factor estimations and constraints based on those~\cite{2022arXiv220710378H}. Moreover, it was recently claimed for Sagittarius~I that a gamma-ray emitting population of millisecond pulsars may be present in dSphs~\cite{2022arXiv220412054C}. This would render the identification of a DM signal in these targets more challenging than long thought.

\section{Dark Subhalos\label{sec:subhalos}}

Not all of the cosmic DM structures may have been massive enough to attract sufficient baryonic matter to form stellar systems. Numerical simulations predict a huge population of DM halos on scales lighter than those of dwarf galaxies, down to the order of Earth mass ($10^{-6}\,\Msol$) and lighter~\cite{2019Galax...7...81Z}. Those halos are called ``dark'', as no baryonic interactions occur inside them which could reveal the positions of DM halos by electromagnetic radiation, except the emission from DM interactions itself. While the absolute rate of relic annihilations or decays in these light objects is lower, they may be found close to the Milky-Way disk and therefore even outshine dSph galaxies in their gamma-ray flux~\cite{2016JCAP...09..047H,2019Galax...7...60H}. However, the existence of such low-mass DM halo population is hypothetical: attempts are done to identify those objects gravitationally on statistical and individual basis by lensing signatures~\cite{2010AdAst2010E...9Z} or stellar tidal streams~\cite{2002MNRAS.332..915I,2009ApJ...705L.223C,2012ApJ...748...20C,2016MNRAS.463..102E,2018JCAP...07..061B,2021JCAP...10..043B}.\endnote{Additionally, a reverse approach was undertaken to search for faint dwarf galaxies with optical telescopes towards the sky directions of \fermi{} unidentified gamma-ray sources~\cite{2018MNRAS.480.2284C}.} Moreover, their mass and spatial distribution in the Galaxy is uncertain, largely affecting the gamma-ray detection prospects at Earth~\cite{2016JCAP...09..047H,2019Galax...7...60H,2021MNRAS.501.3558G}.

One approach of dark subhalo searches is to identify them by their unique gamma-ray signature, which is expected to show (i) no time variability, (ii) a characteristic spectral shape with cut-off according to Figure \ref{fig:dNdE_spectra}, and (iii) a spatial extension with typical emission profile. Dozens of candidates among unidentified sources in the \fermi{}~\cite{2010PhRvD..82f3501B,2012MNRAS.424L..64M,2012A&A...538A..93Z,2012JCAP...11..050Z,2012ApJ...747..121A,2012PhRvD..86d3504B,2014PhRvD..89a6014B,2015JCAP...12..035B,2016JCAP...05..049B,2016ApJ...825...69M,2017PhRvD..95j2001X,2019JCAP...11..045C,2019JCAP...07..020C} and HAWC~\cite{2019JCAP...07..022A} source catalogs have been scrutinized so far without a conclusive result.

\begin{table}[b] 
\caption{Current identification status of dark subhalo candidates so far observed by IACTs among \fermi{} previously unidentified gamma-ray sources. The likely non-DM association of 3FHL J2104.5+2117 was already noted by authors of~\cite{2021ApJ...918...17A}.\label{tab:iact-ufo-status}}
\newcolumntype{A}[1]{>{\centering\arraybackslash}m{#1}}
\begin{tabularx}{\textwidth}{A{3cm}A{3.8cm}A{2.5cm}A{3.8cm}}
\toprule
\textbf{Source Name}	& \textbf{Association} & \textbf{Type}	& \textbf{Reference}\\
\midrule
1FGL J2347.3+0710  & TXS 2344+068 & BL Lac & \cite{2014ApJS..215...14D,2017ApJ...851..135P,2021ApJ...907...67A}  \\
1FGL J0338.8+1313  & RX J0338.4+1302 & BL Lac & \cite{2014ApJS..215...14D,2017ApJ...851..135P,2019AA...632A..77C} \\
2FGL J0545.6+6018 & \multicolumn{3}{c}{\textit{---still unidentified---}} \\
2FGL J1115.0{$-$}0701 & FIRST J111511.6{$-$}070241 & BL Lac &\cite{2013MNRAS.432.1294P,2015ApJ...810...14A,2017AA...602A..86L}\\ 
3FHL J0929.2{$-$}4110&  \multicolumn{3}{c}{\textit{---still unidentified---}} \\
3FHL J1915.2{$-$}1323 &  \multicolumn{3}{c}{\textit{---still unidentified---}} \\
3FHL J2030.2{$-$}5037 & 3HSP J203024.0{$-$}503413  & Possible Blazar & \cite{2019AA...632A..77C} \\ 
3FHL J2104.5+2117 & 3HSP J210415.9+211808  & BL Lac & \cite{2017AA...602A..86L,2017AA...598A..17C,2019AA...632A..77C,2020ApJS..247...33A} \\ 
\bottomrule
\end{tabularx}
\end{table}

Further information about the nature of unidentified gamma-ray sources can be obtained by follow-up observations with IACTs, with higher angular resolution and extending the energy range to search for TeV DM candidates. With respect to \fermi{} gamma-ray sources unassociated at that time, so far eight promising DM subhalo candidates have been subsequently observed by IACTs with the purpose of clarifying their possible DM origin: in 2010, the sources 1FGL J2347.3+0710 and 1FGL J0338.8+1313 have been observed by MAGIC~\cite{2011arXiv1109.5935N}, and between 2013 and 2015, the sources 2FGL J0545.6+6018 and 2FGL J1115.0$-$
0701 by VERITAS~\cite{2015arXiv150900085N}. In 2018 and 2019, \hess{} observed the four sources 3FHL J0929.2{$-$}4110, 3FHL J1915.2{$-$}1323, 3FHL J2030.2{$-$}5037, and 3FHL J2104.5+2117~\cite{2021ApJ...918...17A}. None of the observations showed a signal in very high energy ($E>100\,$GeV) gamma rays. Due to no association with some gravitationally presumed DM budget, no individual constraints on the DM particle physics can be derived from null detections in such searches. However, reversely, under the assumption of some DM particle physics, constraints on the $J$-factor along the corresponding line of sight can be obtained~\cite{2021ApJ...918...17A}. Since their discovery, several of the GeV unassociated gamma-ray sources now have been classified to be probably not of DM origin, as summarized in Table~\ref{tab:iact-ufo-status}. Notably, the nature of three of these sources is still elusive. In addition, follow-up observations of unassociated sources in the HAWC 2HWC catalog have been done by MAGIC~\cite{2019MNRAS.485..356A} and VERITAS~\cite{2018ApJ...866...24A}, though without significant detection and DM interpretation.

A second approach is to head for serendipitous discoveries of dark subhalos in the field of  view of IACTs. While no DM subhalo candidate has been detected with IACTs so far, projections have been made for discovery with the HAWC observatory~\cite{2019JCAP...07..022A} and the future Cherenkov Telescope Array (CTA) during large-scale surveys~\cite{2011PhRvD..83a5003B,2016JCAP...09..047H,2021PDU....3200845C} or serendipitously in the field of view of other targets~\cite{2021PDU....3200845C}.

Third and last, current non-detection or remaining dark subhalo candidates in gamma-ray observations can be translated into constraints on the DM annihilation cross section or lifetime, through assumptions on the global population of DM subhalos in the Galactic halo and their $J$-factors. Such limits on the annihilation cross section have been derived multiple times for the \fermi{} detector~\cite{2010PhRvD..82f3501B,2014PhRvD..89a6014B,2015JCAP...12..035B,2016JCAP...05..028S,2016ApJ...825...69M,2017PhRvD..95f3531L,2019JCAP...11..045C,2019JCAP...07..020C,2017PhRvD..96f3009C,2019Galax...7...90C,2017JCAP...04..018H} and also \mbox{HAWC~\cite{2019JCAP...07..022A,2020Galax...8....5C}.} Analogous constraints from non-detection of DM subhalos in the gamma-ray data of IACTs are yet to be done.

\section{Irregular and Spiral Local Galaxies\label{sec:galaxies}}

Indirect DM searches in individual galaxies other than Galactic dSphs have recently gained increased interest. In particular, this concerns local dwarf Irregular (dIrr) galaxies: these are dwarf disc galaxies rich in gas without a defined disk shape, and showing a relatively low rate of star formation that occurs outside of well-defined spiral arms~\cite{2017MNRAS.465.4703K}. In fact, this type of dwarf galaxies dominates  the Local volume~\cite{2015MNRAS.454.1798K}: while most galaxies in the Local group are dSphs, around twice as many dIrr galaxies than dSphs are expected within a distance of $\sim $$10~$Mpc~\cite{2018PhRvD..98h3008G}, and around 500 dIrrs have already been identified within that sphere~\cite{2013AJ....145..101K,2017MNRAS.465.4703K}. The analysis of 36 spectroscopically well covered dIrrs of distances between 520 kpc and 16.6 Mpc revealed that their DM density profiles can be determined more robustly than for dSphs, and shows average masses of $\sim$$10^{10}\,\Msol$, larger than those of most studied dSphs~\cite{2017MNRAS.465.4703K,2018PhRvD..98h3008G}. Due to their large distances and a cored DM profile preferred by kinematic data, the authors of \cite{2017MNRAS.465.4703K,2018PhRvD..98h3008G} have found annihilation $J$-factors for dIrrs generally 10 to \mbox{100 times} fainter than the brightest dwarfs (see Figure \ref{fig:jfactors},~top panel). However, thanks to the large masses of these dIrrs, an annihilation boost from substructure may significantly increase the signal up to more than one order of magnitude in some cases~\cite{2021PhRvD.104h3026G} (Figure \ref{fig:jfactors}, top panel, light orange sample). Finally, the  cored inner DM profile of dIrrs is still a matter of debate~\cite{2021PhRvD.104h3026G}. Notably, in contrast to dSphs, dIrrs are in principle expected to show an intrinsic gamma-ray background from star-forming regions. However, Ref. \cite{2018PhRvD..98h3008G} has shown that this astrophysical background is negligible for most objects.\endnote{Exceptions are the Large and Small Magellanic Clouds, from which gamma-ray emission has already been detected~\cite{1992ApJ...400L..67S,2010A&A...523A..46A}. Note that the classification of Magellanic spirals like the Large Magellanic Cloud as dIrr galaxies is ambiguous~\cite{2005ApJ...632..872R}.}

At GeV energies, a combined analysis for DM annihilation signals from the direction of 7~dIrr objects in the data from  \fermi{} has been pursued~\cite{2021PhRvD.104h3026G}. Moreover, the Large Magellanic Cloud (LMC)~\cite{2015PhRvD..91j2001B}, Small Magellanic Cloud (SMC)~\cite{2016PhRvD..93f2004C}, and Smith High-Velocity Cloud~\cite{2014ApJ...790...24D} were studied. No DM signal from any of these targets was discovered, with constraints about a factor 10 to several hundreds weaker than those from stacked dSph galaxy data.

These searches have been complemented by recent searches in the TeV regime: H.E.S.S. has observed the Wolf--Lundmark--Melotte (WLM) dIrr galaxy~\cite{2021PhRvD.103j2002A} in the Southern hemisphere (see Figure \ref{fig:skymap_iact}) for 18 h, constraining an annihilation cross section in the order of magnitude $\sigmav{}\lesssim 10^{-21}\,\mathrm{cm^3\,s^{-1}}$ in the range of DM masses of a few TeV. A preliminary stacked analysis by HAWC~\cite{2019ICRC...36..520H} on both DM annihilation and decay in sources in the HAWC field of view (most of them clustered in the upper left of Figure \ref{fig:skymap_fermihawc}) out of the sample from~\cite{2017MNRAS.465.4703K} observed for 1017 days provides annihilation constraints similar to those by H.E.S.S. on WLM, in the same energy range.

In addition, remote spiral galaxies like our own Galaxy, providing classical evidence for the existence of DM~\cite{1939LicOB..19...41B,1983SciAm.248f..96R,2001ARA&A..39..137S}, serve as indirect DM targets. Spiral galaxies consist of a flat rotating disk with a central bulge and contain usually  $10^{10}$ to $\lesssim$$10^{12}$ stars, with DM halos of masses between $\lesssim$$10^{11}$ to $\gtrsim$$10^{12}\,\Msol$, about a factor 5 to 10 heavier than the galaxies' stellar masses~\cite{1992A&AS...92..583F}. However, only the closest spiral galaxies are of interest for DM searches with gamma rays, among them most prominently the M31 (Andromeda) and M33 (Triangulum) galaxies. These two galaxies have been studied using \fermi{} by~\cite{2016JCAP...12..028L,2019PhRvD..99l3027D,2021PhRvD.103b3027K} for M31 and by~\cite{2019PhRvD..99l3027D} for M33. M33 (together with M32) was also searched for DM annihilation before by Whipple~\cite{2008ApJ...678..594W}, and, as one of the very first DM searches with IACTs, the HEGRA data on M31 was analyzed for line signals from DM annihilation~\cite{2003A&A...400..153A}. Lately, M31 was searched for DM at multi-TeV energies using HAWC~\cite{2018JCAP...06..043A}. Despite the very large $J$-factor associated with M31 (see Figure \ref{fig:jfactors}), DM searches in M31 are intricate. Extended gamma-ray emission from the inner 5 kpc (corresponding to $0.4^\circ$ in radius) of M31~\cite{2010A&A...523L...2A,2017ApJ...836..208A} has already been detected by \fermi{}. Interestingly, similar to our own Galaxy, some excess above the expected astrophysical emission has been claimed~\cite{2017ApJ...836..208A}, leading to the discussion of DM interpretations~\cite{2018ApJ...862...79E,2019ApJ...871L...8F,2021PhRvD.103f3023B,2022arXiv220400636Z}. Moreover, like in the Milky Way, gamma radiation can occur in star-forming regions of the M31 disk or Fermi-bubble-like structures, visible under distances up to $\sim$$1^\circ$ from the center of M31. In fact, a weak gamma-ray signal in the \fermi{} data has been claimed up to tens of degrees from the center of M31~\cite{2019ApJ...880...95K}, leading as well to DM interpretations~\cite{2021PhRvD.103b3027K,2022arXiv220512291R}.

All these caveats lead to gamma-ray constraints on DM annihilation on the TeV scale in M31, and similarly in M33, a factor 2 to 10 weaker than the best limits from dSph observations (also for HAWC, which does not see background gamma-ray emission). However, this picture changes when searching for DM decay: here, as shown in Figure \ref{fig:bestlimits_dec}, the best current limit on decaying DM with masses above 10~TeV is obtained by HAWC data on M31. This can be understood by the large M31 $J_{\rm{dec}}$-factor and the, compared to annihilation, less decreased flux from TeV DM decay, as described in Section~\ref{sec:indirect_foundations}. 

Besides M31 and M33, many more spirals are present in the Local volume, although at much larger distances than M31 and M33. An analysis on four low-brightness galaxies (three of them classified as spirals, one as dIrr) resulted in limits about 4 orders of magnitude weaker than from stacked dSphs~\cite{2021MNRAS.501.4238B}. However, recently, Ref.  \cite{2021arXiv210908832H} found that stacking a data set of the order of $10^5$ low-brightness galaxies---which could be readily available with LSST---could provide constraints competitive with other targets. Despite the very early searches with the Whipple telescope, no spiral galaxies have been searched for DM with the current generation of IACTs.

\section{Galaxy Clusters\label{sec:clusters}}

Galaxy clusters form the largest overdensity structures in the Universe, with masses up to more than $10^{15}\,\Msol$, out of which  $\sim$$80\%$ is composed of DM~\cite{2011ARA&A..49..409A}. The average DM density within the central hundreds of kiloparsecs of galaxy clusters is still tens of thousands times higher than the mean Universe's DM density.\endnote{This corresponds to the densities within the so-called characteristic scale radius, $r_{\mathrm{s}}$. Usually, the total extension of a DM halo is given by $R_{200}$, defined as the radius within which the mean density is 200 times higher than the critical density, or about \mbox{600 times} higher than the mean DM density, $\rho_{\mathrm{m}}$. For common clusters, the value of $R_{200}$ is several megaparsecs. For comparison, the average DM density in the inner $r_{\mathrm{s}}\approx 20~$kpc of the Milky Way is about $(1\pm0.2)\times 10^5\,\rho_{\mathrm{m}}$~\cite{2013JCAP...07..016N,2019A&A...621A..56P}.} This makes nearby cluster structures well suited DM targets.

The search for DM in galaxy clusters with gamma rays is generally hampered twofold: first, being located at cosmological distances, absorption at the extragalactic background light comes significantly into play (see Equations~(\ref{eq:flux_1halo_ann})~and~(\ref{eq:flux_1halo_dec})). This attenuates the detectable radiation from these far-away objects in addition to the inverse square-distance law. Second, a variety of other physical processes generating high-energy gamma rays in clusters forms a non-negligible background for DM searches. While, so far, no diffuse gamma rays from cosmic-ray interactions in the inter-cluster medium of galaxy clusters have been clearly detected either by \fermi{} in the GeV~\cite{2015ApJ...812..159A,2016ApJ...819..149A,2021A&A...648A..60A} or by ground-based instruments in the TeV regime~\cite{2009A&A...502..437A,2012ApJ...757..123A,2016A&A...589A..33A}, many clusters host gamma-ray sources such as active galactic nuclei. As blazars, these may form a bright gamma-ray background, with even multiple blazars being present within a single cluster. For DM annihilation, additionally, the detection prospects in clusters strongly depend on the signal boost from the hierarchical DM substructure~\cite{2017MNRAS.466.4974M,2019Galax...7...68A}. The $J_{\rm ann}$-factors (Equation~(\ref{eq:flux_1halo_ann})) of the average DM mass profiles of galaxy clusters are generally somewhat smaller than those of nearby dwarf galaxies. However, multiple levels of substructure are present in galaxy clusters, from the DM halos of their member galaxies, galaxy subhalos, and dark halos down to subhalo masses of $10^{-6}\,\Msol$~\cite{2020Natur.585...39W}. As illustrated in Figure \ref{fig:jfactors}~(top), this may boost the gamma-ray emission by factors of several tens, resulting in $J_{\rm ann}$-factors competitive with those of Galactic dSphs~\cite{2011JCAP...12..011S}. In Figure \ref{fig:bestlimits_ann}, we show the limits from \hess{} observations of the Fornax cluster~\cite{2012ApJ...750..123A}, where a boost of the emission by factor 100 over the considered integration area was assumed compared to the emission from a cluster halo without substructure. It can clearly be seen that such boost  results in constraints similar to those from dSph observations.  In the case of decay, a signal boost from  substructure  is not present. However, due to the large cluster masses seen under a small integration angle, nearby clusters can outplay Galactic dSph galaxies for DM decay searches~\cite{2012PhRvD..86h3506C}, as illustrated in Figure \ref{fig:jfactors}~(bottom) by the $J_{\rm dec}$-factors.

\begin{figure}[t] 
\includegraphics[width= \linewidth]{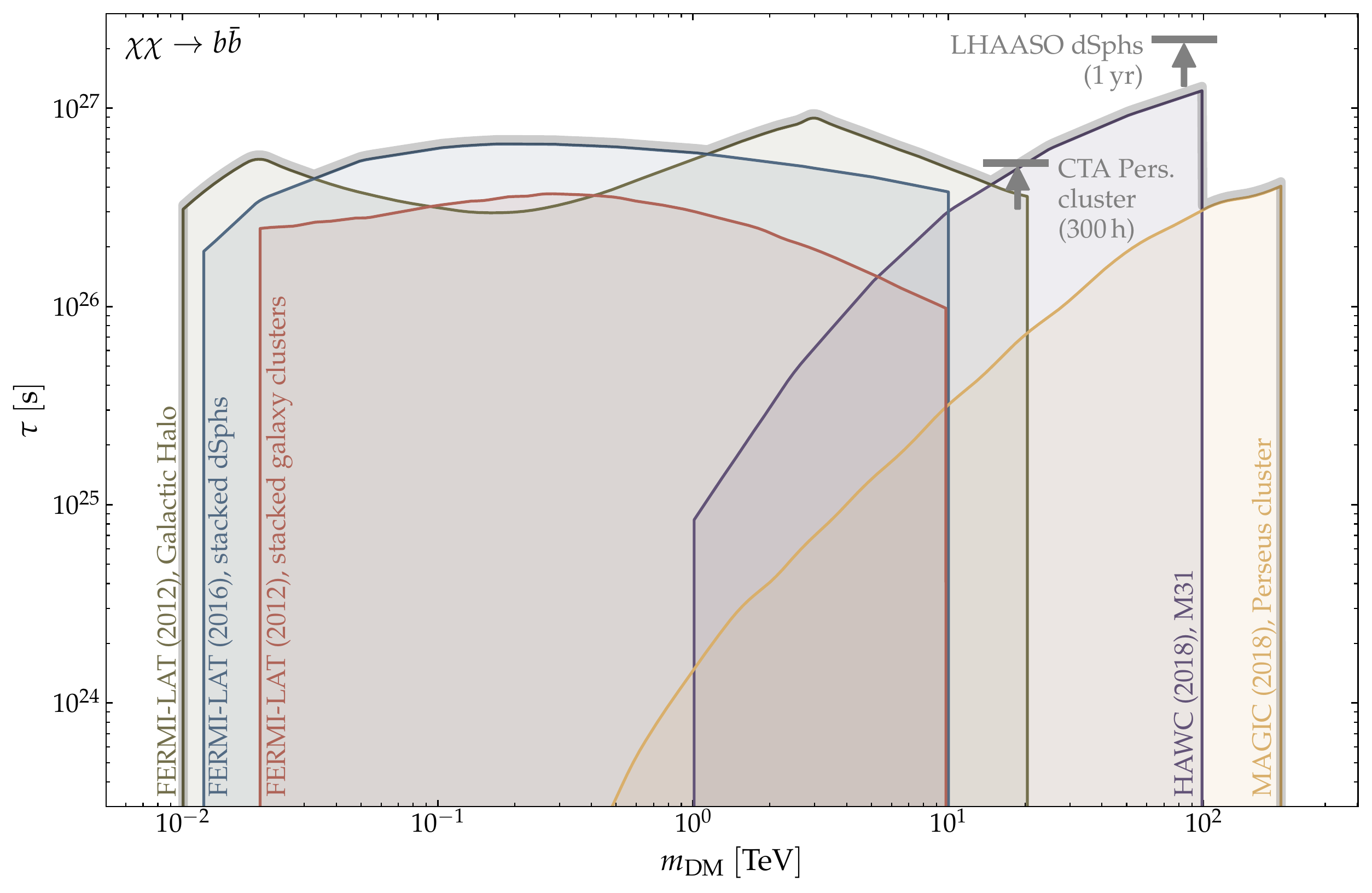}
\caption{The power of searches for decaying TeV DM with ground-based instruments in the extragalactic sky, shown for decays into $b\bar{b}$ quarks: in the lower energy regime, the limits on WIMP lifetime are dominated by \fermi{}'s data on the Galactic halo~\cite{2012ApJ...761...91A}, stacked dwarf spheroidal galaxies~\cite{2016PhRvD..93j3009B}, and stacked galaxy clusters~\cite{2012JCAP...01..042H} (from left to right). For DM masses larger than about 10~TeV, however, constraints are so far only obtained by ground-based instruments from observations of the Perseus cluster by MAGIC~\cite{2018PDU....22...38A} or M31 by HAWC~\cite{2018JCAP...06..043A}. Gray arrows indicate the projected sensitivity reach of future instruments at their most sensitive point in the $m_{\rm DM}$ parameter space for 300$\,$h observations of the Perseus galaxy cluster with CTA~\cite{2019scta.book.....C} and 1 year of combined dSph observations with LHAASO~\cite{2020ChPhC..44h5001H}.}
\label{fig:bestlimits_dec}
\end{figure}

Within an observation campaign covering multiple science purposes, the MAGIC telescopes have observed the Perseus galaxy cluster for almost 400~h~\cite{2018PDU....22...38A}. As shown in Figure \ref{fig:bestlimits_dec}, using more than 200$\,$h of  good-quality data from this data set allowed to present one of the most stringent lower limits as of today on the lifetime of heavy particle DM with masses above 20 TeV. Based on less data, MAGIC~\cite{2010ApJ...710..634A} (also from Perseus), as well as Whipple~\cite{2006ApJ...644..148P} (from Perseus and Abell~2029), VERITAS\endnote{While VERITAS~\cite{2012ApJ...757..123A} published a joint analysis of Coma together with \fermi{} data, their DM limits are only given for the VERITAS observations alone.} \cite{2012ApJ...757..123A} (from the Coma cluster), and, as mentioned already above, \hess{}~\cite{2012ApJ...750..123A} (from the Fornax cluster) derived limits also on DM annihilation from cluster observations.

Many analyses have been performed so far with \fermi{}~\cite{2010JCAP...05..025A,2010JCAP...12..015D,2012MNRAS.427.1651H,2012JCAP...11..048H,2012JCAP...01..042H,2012JCAP...07..017A,2012PhRvD..86g6004M,2013ApJ...762L..22H,2015ApJ...812..159A,2016JCAP...02..026A,2016PhRvD..93j3525L,2018PhRvL.120j1101L,2021MNRAS.502.4039T}. Facilitated by \fermi{}'s all-sky coverage, stacked analyses of many objects were performed, for continuum annihilation~\cite{2012JCAP...07..017A,2012JCAP...01..042H,2018PhRvL.120j1101L,2021MNRAS.502.4039T} and decay~\cite{2010JCAP...12..015D,2012JCAP...01..042H} spectra, and  for line emission from annihilation~\cite{2012JCAP...11..048H,2013ApJ...762L..22H,2016PhRvD..93j3525L,2016JCAP...02..026A,2021MNRAS.502.4039T}. So far, HAWC has published a preliminary analysis for DM annihilation in the Virgo cluster~\cite{2017AIPC.1792f0010C}. While most stacked \fermi{} analyses focused on a selections of dozens or less most promising targets, others included hundreds~\cite{2018PhRvL.120j1101L} or thousands~\cite{2016arXiv160609642Q} of objects from galaxy catalogs.\endnote{Note that in these cases of very large samples, sky regions of cluster structures and other closely located objects overlap, as e.g., individual local dIrr galaxies in~\cite{2018PhRvL.120j1101L}.} Stacking of large target samples is also done in searches for cross-correlation between gamma rays and galaxy catalogs~\cite{2015PhRvL.114x1301R,2015ApJS..221...29C,2016JCAP...06..045A,2020MNRAS.495..114T}, gravitational tracers~\cite{2014PhRvD..90f3502S,2016PhRvD..94f3522S,2017MNRAS.467.2706T} or constrained simulations of the large-scale structure~\cite{2022arXiv220512916B}. Additionally, without correlating to spatial emission structures, the diffuse gamma-ray background was searched for annihilation~\cite{2015PhRvD..91l3001D} or decay~\cite{2010JCAP...06..027Z,2012JCAP...01..042H,2015JCAP...05..024A}, as well as the diffuse background's angular power spectrum~\cite{2012PhRvD..85h3007A,2016PhRvD..94l3005F}. In  Figure \ref{fig:skymap_fermihawc}, we have included the 18~individual galaxy cluster targets studied multiple times for DM signals so far. Particularly for searches for DM decay, deep exposures on galaxy clusters by TeV instruments have turned out to be very powerful, as shown in Figure \ref{fig:bestlimits_dec} (yellow curve) for the result of MAGIC towards the Perseus cluster of galaxies~\cite{2018PDU....22...38A}. Such competitive limits compared to measurements at lower energies are  possible because, as mentioned previously, the gamma-ray flux after DM decays does not drop as fast with the particle mass as for annihilation. Moreover, the angular emission profile from decay is much broader than the one from annihilation, and can be resolved, as in the case of Perseus, by the high angular resolution of IACTs. This  allows to separate regions where only emission from DM is expected from conflicting astrophysical emitters, as e.g., the active galactic nucleus NCG~1275 in the Perseus cluster.

\section{Multi-Wavelength, Multi-Messenger, and Multi-Instrument Combined Analyses\label{sec:mwl}}

In recent years, there have been increasingly growing efforts in expanding the reach of DM searches with gamma rays. These expansions are following three main avenues: multi-instrument, multi-wavelength, and  multi-messenger analyses. The first two already started with the first joint analysis by \fermi{} and MAGIC published in 2016~\cite{2016JCAP...02..039M} and cover a wide range of masses from 10~GeV to 100~TeV. Since then, efforts have been expanded to a joint analysis by five gamma-ray detectors, namely, the gamma-ray satellite \fermi{}, the water Cherenkov observatory HAWC, and the IACTs H.E.S.S., MAGIC, and VERITAS. These efforts largely resulted in a standardization of the analyses developed in each collaboration and a consistent analysis framework. A preliminary combined analysis on DM masses from 5~GeV to~100 TeV could already exclude a thermal relic cross section of DM masses up to $\sim$200~GeV in the most optimistic scenario~\cite{2019ICRC...36...12O,2022icrc.confE.528A}. In the future, such joint searches can easily be extended to other gamma-ray detectors such as LHAASO which shows excellent capabilities to DM searches in the 1 TeV to 1 PeV range~\cite{2019PhRvD.100h3003H,2020ChPhC..44h5001H}.

Beyond these experiments covering the GeV to PeV range, DM searches are also done using photons at lower energies, while using the same or similar targets, see \mbox{e.g.,~\cite{2013ApJ...773...61S,2013PhRvD..88h3535N,2016PhRvD..94b3507C,2017JCAP...07..025R,2018PhRvD..97j3021M,2019PhRvD.100d3002K,2020MNRAS.496.2663V,2020PhRvD.102f3017C,2022JHEP...04..018Y,2022PhRvD.105l3006C}.} For example, very strong constraints have been recently obtained on GeV to TeV DM annihilation from radio data on the LMC~\cite{2021JCAP...11..046R}. There is nothing preventing the effort of extending a combined and harmonized analysis to such wavelengths. Finally, the neutrino detectors ANTARES and IceCube recently reported on a combined analysis of their data searching for DM in the direction of the GC~\cite{2020PhRvD.102h2002A}, and also a combination of DM searches between gamma rays and neutrinos would be promising. For further details on multi-messenger TeV DM searches, we refer the reader to the review by~\cite{2019FrASS...6...19G}.

\section{Primordial Black Hole Evaporation\label{sec:pbhs}}

At the end of Section~\ref{sec:subhalos}, we have remarked that a blind search for gamma-ray signals from DM subhalos in the accumulated data sets of IACTs is yet to be done. However, archival ``blind'' searches in large IACT data sets have already been pursued for a DM candidate unrelated to WIMPs, namely, PBHs. This shows that gamma-ray data sets can be used to also constrain a completely different class of DM candidates. PBHs are thought to have formed in the radiation-dominated early Universe, and thus do not belong to the baryonic mass budget inferred by CMB measurements~\cite{2020ARNPS..70..355C,2021RPPh...84k6902C,2021arXiv211002821C}. Therefore, they would be viable candidates to contribute to the cosmic DM budget. They can be searched with many different methods among which gravitational lensing, measurements of dynamical gravitational distortions like in stellar streams~\cite{2022PDU....3500978M} or emission of gravitational waves~\cite{2021JPhG...48d3001G}.

In particular, lighter PBHs can be probed with the gamma rays from PBHs evaporating via Hawking radiation: any black holes lose mass, i.e., evaporate, via Hawking radiation. From the Hawking radiation into leptons, quarks, and gauge bosons, secondary gamma rays are produced traveling away from the evaporating black hole~\cite{2016APh....80...90U}. The closer a PBH comes to the end of its life, the stronger and energetic the gamma-ray emission  becomes~\cite{2016APh....80...90U,2021JCAP...12..051C}. Specifically, the PBH lifetime is given by:
\begin{equation}
    t(M) \sim \frac{G^2 M^3}{\hbar c^4} \sim 10^{64} \left( \frac{M}{M_{\odot}}\right)^3 \,\mathrm{yr}
\end{equation}
where $M$ is the mass of the PBH. Consequently, only PBHs created in the early Universe with masses heavier than $\sim$$5 \times 10^{14}\,\mathrm{g} = 2.5\times 10^{-19}\,\Msol$ have survived until today. This allowed for strong constraints to PBH masses slightly above masses of $\sim$$5 \times 10^{14}\,\mathrm{g}$ from measurements of the isotropic gamma-ray background~\cite{2021JPhG...48d3001G}. At the very end of the PBHs life, the gamma-ray flux light curve of the evaporation is expected to increase following a power law, resulting in a sudden burst  of energy in the last seconds. The gamma-ray spectrum of this final-stage emission extends up to TeV energies, and can be detected by gamma-ray instruments as sudden TeV gamma-ray bursts. However, there are two crucial differences to known gamma-ray bursts of cosmological origin:\endnote{See~\cite{2016APh....80...90U} for a complete list of differences between cosmological  and PBH gamma-ray bursts.} First, unlike gamma-ray bursts from collapse or merger of compact extragalactic objects, detectable gamma-ray flashes from evaporating PBHs originate only from nearby PBHs of distance within several parsecs and therefore, are not attenuated by absorption from the extragalactic background light. Secondly, the duration of the peak PBH burst emission is very short, happening on timescales of seconds~\cite{2016APh....80...90U}. Unlike in cosmological gamma-ray bursts, no afterglow emission is expected from PBHs, and follow-up observations  with pointed telescopes of PBH evaporation candidates seen by gamma-ray burst monitors are therefore virtually impossible. As a consequence, only the archival search for serendipitously recorded burst events in the data of IACTs and air shower arrays is possible. In addition, while measurements of the isotropic gamma-ray background have  limited PBHs of masses $\gtrsim$ $5 \times 10^{14}\,\mathrm{g}$ to only contribute less than a fraction of $\lesssim$$10^{-6}$ to the cosmic DM budget, PBHs of this mass scale could have accumulated in galactic halos, resulting in a significantly higher burst rate detectable in their galactic neighborhood. While current limits correspond to a cosmic average PBH burst rate of $\lesssim$$10^{-6}$ bursts pc$^{-3}$ yr$^{-1}$, in the case of galactic PBH clusters, these limits correspond to about one burst pc$^{-3}$ yr$^{-1}$ in the Galactic neighborhood according to~\cite{1976ApJ...206....1P}. Such a rate is in reach to be probed by gamma-ray detection of individual bursts. However, the search for individual burst events would be independent from isotropic measurements and from the assumption of a given clustering scale~\cite{2016APh....80...90U}.

IACTs have been used to set constraints on the evaporation rate since Whipple~\cite{2006JCAP...01..013L} reached an upper limit of $1.08 \times 10^{6}$ pc$^{-3}$ yr$^{-1}$ (99\% CL) on the evaporation rate in the Solar neighborhood (several parsecs) in the Galaxy, similar to the constraints found by the CYGNUS~\cite{1993PhRvL..71.2524A} and TIBET~\cite{1996A&A...311..919T} air shower arrays. Since then, H.E.S.S.~\cite{2013ICRC...33.2729G} (2013), \cite{2019ICRC...36..804T} (2019) and~\cite{2022icrc.confE.518T} (2021) has improved these searches up to constraints of $\lesssim$$10^{3}$ pc$^{-3}$ yr$^{-1}$ for time intervals of the final evaporation between 10 and 100 seconds. VERITAS~\cite{2012JPhCS.375e2024T} (2012), \cite{2017ICRC...35..691A} (2017) derived similar preliminary limits to a burst rate of $\leq$$2.2\times 10^{4}$ pc$^{-3}$ yr$^{-1}$ for a search window of 30$\,$s using 747$\,$h of data~\cite{2019ICRC...36..719K}, with ongoing analyses to improve that result~\cite{2019ICRC...36..719K,2022icrc.confE.822P}. Limits have also been derived by Milagro~\cite{2015APh....64....4A}, \fermi{}~\cite{2018ApJ...857...49A}, and recently, HAWC~\cite{2020JCAP...04..026A}. A recent study for the Southern Wide-field Gamma-ray Observatory (SWGO) showed how the sensitivity to detect PBH evaporation bursts in TeV gamma rays can be pushed down to rates below 100 bursts pc$^{-3}$ yr$^{-1}$~\cite{2021JCAP...08..040L} in the Solar neighborhood. 
These constraints show how individual PBH evaporation burst events can be competitively searched for with TeV gamma-ray instruments, and that small field of view telescopes like IACTs show capabilities comparable to large field of view detectors like \fermi{} and air shower arrays. Moreover, competitive constraints can be achieved with the MAGIC telescopes and the future CTA~\cite{Cassanyes:2015wpr}.

\newpage 
\section{Future Searches and Conclusions\label{sec:conclusions}}

In this review, we have demonstrated the merits of the many targets in the extragalactic sky to search for TeV WIMP DM and PBHs with gamma rays. In Figure \ref{fig:bestlimits_ann}, the current best constraints on the WIMP annihilation cross section obtained from various targets and (combined) instruments are compiled for a smooth final-state spectrum, e.g., for annihilations into $b\bar{b}$ quarks). While for WIMP masses below $\sim$$1$~TeV, these constraints are mainly dominated by \fermi{}, for higher masses, results from IACTs and water Cherenkov detectors like HAWC become strongly competitive. Moreover, constraints from different targets, or after combining different targets, give comparable results: this illustrates the robustness of the constraints obtained with multiple targets. In the case of searches for decaying WIMPs, observations of extragalactic targets even compete with constraints from the Galactic halo, and  for masses above several TeV, measurements with ground-based instruments compete with space-borne searches as illustrated in Figure \ref{fig:bestlimits_dec}. In addition, high-energy gamma-ray telescopes can contribute to searches for PBHs as potential alternative DM candidate by trawling their (archival) data for short-term gamma-ray bursts from evaporating PBHs in the Solar neighborhood.

In the recent years, the combination of the data from several instruments allowed to further improve the sensitivity for WIMP DM searches with gamma rays and to reduce the uncertainties from possible instrumental systematics. In the latest ongoing work by~\cite{2019ICRC...36...12O,2022icrc.confE.528A}, a joint analysis over the data on 20~dSphs, obtained with five~instruments using three different detection principles is performed. Such analyses may play an even increasingly important role in the future. Besides a variety of targets and different detection techniques of gamma rays, also different messengers like neutrinos can be included in a joint analysis. Such multi-instrument analysis approaches could be the key answer to the quest for heavy particle DM. This is already illustrated by strong constraints from radio signals~\cite{2021Galax...9...11C}, like e.g., towards the LMC~\cite{2021JCAP...11..046R} or Galactic dSph galaxies~\cite{2020MNRAS.494..135C}. In any case, a diverse pool of observations on various targets and with various instruments will be crucial. This also includes dedicated searches for exotic spectral features like lines, ``boxes'', or sharp cut-offs, which may provide a distinctive smoking gun evidence for DM. 

In parallel, next-generation instruments will provide a further boosted sensitivity to probe WIMP annihilations on the mass scale beyond several hundreds of GeV to multiples of TeV: big hopes lie here on CTA, which is currently being built. Consisting of a total of about a 100~IACTs divided into two arrays in the Northern and Southern hemisphere respectively, CTA is predicted to achieve a sensitivity to gamma-ray signals from TeV DM a factor 5 to 10 higher than current IACTs in the same observation time. This sensitivity is facilitated by (i) a decreased energy threshold, allowing to detect more of the low energy photons after WIMP annihilation or decay (see Figure \ref{fig:dNdE_spectra}), (ii) a factor 10 increased effective detection area, and (iii) even better angular resolution than current IACTs. In Figures~\ref{fig:bestlimits_ann}~and~\ref{fig:bestlimits_dec}, we show with the gray arrows how CTA can correspondingly reach a regime beyond the current limits towards dSphs. In Figure \ref{fig:bestlimits_ann}, one can see how the best sensitivity for CTA also shifts to lower energies compared to current IACTs, thanks to its improved low-energy threshold. Besides this, also the air shower array LHAASO~\cite{2019PhRvD.100h3003H,2020ChPhC..44h5001H} and the planned water Cherenkov detector SWGO~\cite{2022icrc.confE.555V} promise improved sensitivity to WIMP masses at beyond-TeV energies; in Figures~\ref{fig:bestlimits_ann}~and~\ref{fig:bestlimits_dec}, it is shown how LHAASO may provide unprecedented constraints from dSph targets beyond 10~TeV.
   
In conclusion, indirect searches for various DM candidates with high-energy gamma rays constitute a central building block in ongoing endeavor to unveil the nature of DM. In particular, indirect detection of TeV WIMP DM in gamma rays from multiple targets in the extragalactic sky will provide a crucial piece of information for a clear DM identification: the signal will be seen from the same objects that show gravitational evidence for DM, and with a universal signature from different targets, many of them free of confounding backgrounds.   With this, we advocate these searches to be continued in the future, with novel detectors and improved analysis techniques. To this end, we hope that this review, and in particular Figures~\ref{fig:skymap_iact} and \ref{fig:skymap_fermihawc} and Table~\ref{tab:fermihawc-targetsref}, may provide a clear guidance to which targets and searches are yet to be complemented with improved methods and observations.

\vspace{6pt} 

\authorcontributions{Initial draft---both authors. Writing---Sections~\ref{sec:intro}, \ref{sec:indirect_foundations}, \ref{sec:subhalos}, and  \ref{sec:clusters}: M.H., Sections~\ref{sec:dwarfs} and \ref{sec:mwl}: D.K., Sections~\ref{sec:galaxies}, \ref{sec:pbhs}, and \ref{sec:conclusions}: both authors. Figures~\ref{fig:skymap_iact}, \ref{fig:skymap_fermihawc}, \ref{fig:bestlimits_ann}, and  \ref{fig:bestlimits_dec}: M.H., Figures~\ref{fig:jfactors} and \ref{fig:dNdE_spectra}: both authors. Table~\ref{tab:fermihawc-targetsref}: M.H. All authors have read and agreed to the published version of the manuscript.}

\funding{M.H. acknowledges funding from the University of Tokyo and the ICRR Inter-University Research Program in the Fiscal Years 2021 and 2022. D.K. acknowledges funding from the ERDF under the Spanish Ministerio de Ciencia e Innovaci\'{o}n (MICINN), grant PID2019-107847RB-C41, and from the CERCA program of the Generalitat de~Catalunya.}

\dataavailability{Data used in the figures of this review are available upon request.}

\acknowledgments{The authors would like to thank the  {Special Issue Editors John Quinn, Deirdre Horan, and Elisa Pueschel} 
 for the invitation to compile this review. Moreover, the authors would like to thank Javier Rico and Francesco G. Saturni
 for providing useful comments that helped to substantially improve the quality of the manuscript. This article makes use of \texttt{matplotlib}~\cite{Hunter:2007} and \url{https://carto.com/carto-colors} color schemes provided by \url{https://jiffyclub.github.io/palettable} (websites accessed on July 30, 2022).}

\conflictsofinterest{The authors declare no conflict of interest.}

\abbreviations{Abbreviations}{
The following abbreviations are used in this manuscript:\\

\noindent 
\begin{tabular}{@{}ll}
ALP & axion-like particle\\
ANTARES & Astronomy with a Neutrino Telescope and Abyss environmental RESearch project\\
CTA & Cherenkov Telescope Array\\
CMB & cosmic microwave background\\
DM  & dark matter\\
DOF  & degree of freedom\\
dIrr & dwarf irregular galaxy\\
dSph & dwarf spheroidal galaxy\\
GC   & Galactic center\\
HAWC & High-Altitude Water Cherenkov Gamma-Ray Observatory\\
HEGRA & High Energy Gamma Ray Astronomy (experiment)\\
H.E.S.S.  & High Energy Stereoscopic System\\
IACT & imaging atmospheric Cherenkov telescope\\
LAT  & Large Area Telescope\\
LHAASO & Large High Altitude Air Shower Observatory\\
LMC  & Large Magellanic Cloud\\
LSST & Legacy Survey of Space and Time\\
MAGIC & Major Atmospheric Gamma Imaging Cherenkov telescopes\\
NFW  & Navarro-Frenk-White (DM profile)\\
PBH  & primordial black hole\\
SMC  & Small Magellanic Cloud\\
SWGO & Southern Wide-field Gamma-ray Observatory\\
VERITAS & Very Energetic Radiation Imaging Telescope Array System\\
WIMP & weakly interacting massive particle\\
WLM  & Wolf-Lundmark-Melotte (dIrr galaxy)\\
\end{tabular}}
\newpage
\appendixtitles{no} \appendixstart
\appendix
\section[\appendixname~\thesection]{}
\vspace{-6pt}

\begin{table}[H] 
\caption{Names and references to the DM analyses in \fermi{} and HAWC data of the sources displayed in Figure \ref{fig:skymap_fermihawc}. For galaxy clusters, only sources are listed that are individually analyzed for DM in at least two publications in our literature search. Further clusters are, e.g., studied \mbox{in~\cite{2012JCAP...07..017A,2016PhRvD..93j3525L,2018PhRvL.120j1101L}} and implicitly in cross-correlation or power spectrum studies (see Section~\ref{sec:clusters} for further references.) \label{tab:fermihawc-targetsref}} 
	\begin{adjustwidth}{-\extralength}{0cm}
\newcolumntype{C}{>{\centering\arraybackslash}X}
\begin{tabularx}{\fulllength}{CCC}
\toprule

\textbf{Index in Figure \ref{fig:skymap_fermihawc}}	& \textbf{Source Name}	&  \textbf{Reference}\\
\midrule
\multicolumn{3}{c}{Dwarf spheroidal galaxies}\\\midrule
1 & Aquarius II                & \cite{2020JCAP...02..012H,2018PhRvD..97h3007L} \\
2 & Boötes I                   & \cite{2010ApJ...712..147A,2010JCAP...12..015D,2011PhRvL.107x1302A,2012PhRvD..86b1302G,2012JCAP...11..048H,2012APh....37...26M,2014PhRvD..89d2001A,2015PhRvL.115w1301A,2015ApJS..219...37P,2016MNRAS.461.3976P,2016PhRvD..93j3009B,2016JCAP...02..039M,2017ApJ...834..110A,2018ApJ...853..154A,2018JCAP...03..010P,2018JCAP...10..029C,2019JCAP...04..048Q,2020PhRvD.101j3001A,2020JCAP...02..012H,2016PhRvD..94j3502L,2015JCAP...09..016H,2011PhRvL.107x1303G,2018PhRvD..97h3007L} \\
3 & Boötes II                  & \cite{2010ApJ...712..147A,2014PhRvD..89d2001A,2015PhRvL.115w1301A,2015ApJS..219...37P,2017ApJ...834..110A,2021PhRvD.104h3037L,2019JCAP...04..048Q,2018PhRvD..97l2001L,2016PhRvD..94j3502L,2015JCAP...09..016H,2018PhRvD..97h3007L} \\
4 & Boötes III                 & \cite{2014PhRvD..89d2001A,2015PhRvL.115w1301A,2015ApJS..219...37P,2017ApJ...834..110A,2021PhRvD.104h3037L,2018PhRvD..97l2001L,2016PhRvD..94j3502L,2015JCAP...09..016H,2018PhRvD..97h3007L} \\
5 & Canes Venatici I           & \cite{2014PhRvD..89d2001A,2015PhRvL.115w1301A,2015ApJS..219...37P,2016PhRvD..93j3009B,2017ApJ...834..110A,2018ApJ...853..154A,2018JCAP...10..029C,2020JCAP...02..012H,2020PhRvD.101j3001A,2016PhRvD..94j3502L,2015JCAP...09..016H,2018PhRvD..97h3007L} \\
6 & Canes Venatici II          & \cite{2020PhRvD.101j3001A,2018ApJ...853..154A,2016JCAP...02..039M,2017ApJ...834..110A,2020JCAP...02..012H,2014PhRvD..89d2001A,2015PhRvL.115w1301A,2015ApJS..219...37P,2016PhRvD..93j3009B,2019JCAP...04..048Q,2018JCAP...03..010P,2016MNRAS.461.3976P,2018JCAP...10..029C,2019JCAP...04..048Q,2016PhRvD..94j3502L,2015JCAP...09..016H,2018PhRvD..97h3007L} \\
7 & Carina                     & \cite{2016JCAP...02..039M,2017ApJ...834..110A,2020JCAP...02..012H,2012APh....37...26M,2014PhRvD..89d2001A,2015PhRvL.115w1301A,2011PhRvL.107x1302A,2015ApJS..219...37P,2016PhRvD..93j3009B,2019JCAP...04..048Q,2018JCAP...03..010P,2016MNRAS.461.3976P,2020JCAP...09..004A,2018JCAP...10..029C,2019JCAP...04..048Q,2016PhRvD..94j3502L,2015JCAP...09..016H,2012PhRvD..86b3528C,2018PhRvD..97h3007L} \\
8 & Carina II                  & \cite{2020JCAP...02..012H,2018PhRvD..97h3007L} \\
9 & Coma Berenices             & \cite{2020PhRvD.101j3001A,2018ApJ...853..154A,2016JCAP...02..039M,2017ApJ...834..110A,2020JCAP...02..012H,2012APh....37...26M,2014PhRvD..89d2001A,2015PhRvL.115w1301A,2010ApJ...712..147A,2010JCAP...12..015D,2011PhRvL.107x1302A,2015ApJS..219...37P,2016PhRvD..93h3513Z,2016PhRvD..93j3009B,2019JCAP...04..048Q,2018JCAP...03..010P,2016MNRAS.461.3976P,2021PhRvD.104h3037L,2018JCAP...10..029C,2019JCAP...04..048Q,2018PhRvD..97l2001L,2016PhRvD..94j3502L,2015JCAP...09..016H,2018PhRvD..97h3007L} \\
10 & Draco                      & \cite{2018ApJ...853..154A,2016JCAP...02..039M,2017ApJ...834..110A,2020JCAP...02..012H,2012APh....37...26M,2014PhRvD..89d2001A,2015PhRvL.115w1301A,2010ApJ...712..147A,2010JCAP...12..015D,2011PhRvL.107x1302A,2012PhRvD..86b1302G,2012JCAP...11..048H,2015ApJS..219...37P,2016PhRvD..93h3513Z,2016PhRvD..93j3009B,2019JCAP...04..048Q,2018JCAP...03..010P,2016MNRAS.461.3976P,2018JCAP...10..029C,2020JCAP...09..004A,2016PhRvD..94j3502L,2015JCAP...09..016H,2012PhRvD..86b3528C,2011PhRvL.107x1303G,2018PhRvD..97h3007L} \\
11 & Draco II                   & \cite{2017ApJ...834..110A,2020JCAP...02..012H,2021PhRvD.104h3037L,2018PhRvD..97l2001L,2016PhRvD..94j3502L,2018PhRvD..97h3007L} \\
12 & Fornax                     & \cite{2016JCAP...02..039M,2017ApJ...834..110A,2020JCAP...02..012H,2012APh....37...26M,2014PhRvD..89d2001A,2015PhRvL.115w1301A,2010ApJ...712..147A,2010JCAP...12..015D,2011PhRvL.107x1302A,2012PhRvD..86b1302G,2012JCAP...11..048H,2015ApJS..219...37P,2016PhRvD..93j3009B,2019JCAP...04..048Q,2018JCAP...03..010P,2016MNRAS.461.3976P,2020JCAP...09..004A,2018JCAP...10..029C,2019JCAP...04..048Q,2016PhRvD..94j3502L,2015JCAP...09..016H,2012PhRvD..86b3528C,2011PhRvL.107x1303G,2018PhRvD..97h3007L} \\
13 & Hercules                   & \cite{2020PhRvD.101j3001A,2018ApJ...853..154A,2016JCAP...02..039M,2017ApJ...834..110A,2020JCAP...02..012H,2014PhRvD..89d2001A,2015PhRvL.115w1301A,2010ApJ...712..147A,2015ApJS..219...37P,2016PhRvD..93j3009B,2019JCAP...04..048Q,2018JCAP...03..010P,2016MNRAS.461.3976P,2018JCAP...10..029C,2019JCAP...04..048Q,2016PhRvD..94j3502L,2015JCAP...09..016H,2018PhRvD..97h3007L} \\
14 & Horologium                 & \cite{2017ApJ...834..110A,2018JCAP...10..029C,2020JCAP...02..012H,2016PhRvD..94j3502L,2015JCAP...09..016H,2015ApJ...809L...4D,2018PhRvD..97h3007L} \\
15 & Hydra II                   & \cite{2017ApJ...834..110A,2018JCAP...10..029C,2016PhRvD..94j3502L,2018PhRvD..97h3007L} \\
16 & Leo I                      & \cite{2020PhRvD.101j3001A,2018ApJ...853..154A,2017ApJ...834..110A,2020JCAP...02..012H,2014PhRvD..89d2001A,2015PhRvL.115w1301A,2015ApJS..219...37P,2016PhRvD..93j3009B,2018JCAP...10..029C,2020JCAP...09..004A,2016PhRvD..94j3502L,2015JCAP...09..016H,2012PhRvD..86b3528C,2018PhRvD..97h3007L} \\
17 & Leo II                     & \cite{2020PhRvD.101j3001A,2018ApJ...853..154A,2016JCAP...02..039M,2017ApJ...834..110A,2020JCAP...02..012H,2014PhRvD..89d2001A,2015PhRvL.115w1301A,2015ApJS..219...37P,2016PhRvD..93j3009B,2019JCAP...04..048Q,2018JCAP...03..010P,2016MNRAS.461.3976P,2020JCAP...09..004A,2018JCAP...10..029C,2019JCAP...04..048Q,2016PhRvD..94j3502L,2015JCAP...09..016H,2012PhRvD..86b3528C,2018PhRvD..97h3007L} \\
18 & Leo IV                     & \cite{2020PhRvD.101j3001A,2018ApJ...853..154A,2016JCAP...02..039M,2017ApJ...834..110A,2020JCAP...02..012H,2014PhRvD..89d2001A,2015PhRvL.115w1301A,2010ApJ...712..147A,2015ApJS..219...37P,2016PhRvD..93j3009B,2019JCAP...04..048Q,2018JCAP...03..010P,2016MNRAS.461.3976P,2018JCAP...10..029C,2019JCAP...04..048Q,2016PhRvD..94j3502L,2015JCAP...09..016H,2018PhRvD..97h3007L} \\
19 & Leo V                      & \cite{2017ApJ...834..110A,2020JCAP...02..012H,2014PhRvD..89d2001A,2015PhRvL.115w1301A,2015ApJS..219...37P,2016PhRvD..93j3009B,2018JCAP...10..029C,2016PhRvD..94j3502L,2015JCAP...09..016H,2018PhRvD..97h3007L} \\
20  & Leo T                       & \cite{2016PhRvD..94j3502L,2018PhRvD..97h3007L}\\
21 & Pisces II                  & \cite{2017ApJ...834..110A,2020JCAP...02..012H,2014PhRvD..89d2001A,2015PhRvL.115w1301A,2015ApJS..219...37P,2018JCAP...10..029C,2016PhRvD..94j3502L,2015JCAP...09..016H,2018PhRvD..97h3007L} \\
22 & Reticulum II               & \cite{2016PhRvD..93j3009B,2017ApJ...834..110A,2020JCAP...02..012H,2018JCAP...10..029C,2021PhRvD.104h3037L,2018PhRvD..97l2001L,2018ChPhC..42b5102Z,2016PhRvD..94j3502L,2015JCAP...09..016H,2015PhRvL.115h1101G,2015ApJ...809L...4D,2018PhRvD..97h3007L} \\
23 & Sagittarius~I                & \cite{2014PhRvD..89d2001A,2015PhRvL.115w1301A,2016PhRvD..94j3502L,2015JCAP...09..016H,2018PhRvD..97h3007L} \\
24 & Sculptor                   & \cite{2016JCAP...02..039M,2017ApJ...834..110A,2020JCAP...02..012H,2012APh....37...26M,2014PhRvD..89d2001A,2015PhRvL.115w1301A,2010ApJ...712..147A,2010JCAP...12..015D,2011PhRvL.107x1302A,2012PhRvD..86b1302G,2012JCAP...11..048H,2015ApJS..219...37P,2016PhRvD..93j3009B,2019JCAP...04..048Q,2018JCAP...03..010P,2016MNRAS.461.3976P,2020JCAP...09..004A,2018JCAP...10..029C,2019JCAP...04..048Q,2016PhRvD..94j3502L,2015JCAP...09..016H,2012PhRvD..86b3528C,2011PhRvL.107x1303G,2018PhRvD..97h3007L} \\
25 & Segue 1                    & \cite{2020PhRvD.101j3001A,2018ApJ...853..154A,2016JCAP...02..039M,2017ApJ...834..110A,2020JCAP...02..012H,2012PhRvD..86f3521B,2012APh....37...26M,2014PhRvD..89d2001A,2015PhRvL.115w1301A,2011PhRvL.107x1302A,2012PhRvD..86b1302G,2012JCAP...11..048H,2015ApJS..219...37P,2016PhRvD..93h3513Z,2016PhRvD..93j3009B,2019JCAP...04..048Q,2018JCAP...03..010P,2016MNRAS.461.3976P,2018JCAP...10..029C,2021PhRvD.104h3037L,2020JCAP...09..004A,2019JCAP...04..048Q,2018PhRvD..97l2001L,2016PhRvD..94j3502L,2015JCAP...09..016H,2011PhRvL.107x1303G,2010JCAP...01..031S,2018PhRvD..97h3007L} \\
26 & Segue II                   & \cite{2014PhRvD..89d2001A,2015PhRvL.115w1301A,2010ApJ...712..147A,2015ApJS..219...37P,2016PhRvD..93j3009B,2016PhRvD..94j3502L,2015JCAP...09..016H,2018PhRvD..97h3007L} \\
27 & Sextans                    & \cite{2020PhRvD.101j3001A,2018ApJ...853..154A,2016JCAP...02..039M,2017ApJ...834..110A,2020JCAP...02..012H,2012APh....37...26M,2014PhRvD..89d2001A,2015PhRvL.115w1301A,2010ApJ...712..147A,2010JCAP...12..015D,2011PhRvL.107x1302A,2012PhRvD..86b1302G,2012JCAP...11..048H,2015ApJS..219...37P,2016PhRvD..93j3009B,2019JCAP...04..048Q,2018JCAP...03..010P,2016MNRAS.461.3976P,2018JCAP...10..029C,2020JCAP...09..004A,2019JCAP...04..048Q,2016PhRvD..94j3502L,2015JCAP...09..016H,2012PhRvD..86b3528C,2011PhRvL.107x1303G,2018PhRvD..97h3007L} \\
28 & Tucana II                  & \cite{2017ApJ...834..110A,2020JCAP...02..012H,2018JCAP...10..029C,2019JCAP...08..028B,2016PhRvD..94j3502L,2015JCAP...09..016H,2015ApJ...809L...4D,2018PhRvD..97h3007L} \\
29 & Ursa Major I               & \cite{2018ApJ...853..154A,2017ApJ...834..110A,2020JCAP...02..012H,2014PhRvD..89d2001A,2015PhRvL.115w1301A,2010ApJ...712..147A,2015ApJS..219...37P,2016PhRvD..93j3009B,2018JCAP...10..029C,2016PhRvD..94j3502L,2015JCAP...09..016H,2018PhRvD..97h3007L} \\
30 & Ursa Major II              & \cite{2018ApJ...853..154A,2016JCAP...02..039M,2017ApJ...834..110A,2020JCAP...02..012H,2012APh....37...26M,2014PhRvD..89d2001A,2015PhRvL.115w1301A,2010ApJ...712..147A,2010JCAP...12..015D,2011PhRvL.107x1302A,2015ApJS..219...37P,2016PhRvD..93h3513Z,2016PhRvD..93j3009B,2019JCAP...04..048Q,2018JCAP...03..010P,2016MNRAS.461.3976P,2021PhRvD.104h3037L,2018JCAP...10..029C,2019JCAP...04..048Q,2018PhRvD..97l2001L,2016PhRvD..94j3502L,2015JCAP...09..016H,2018PhRvD..97h3007L} \\
31 & Ursa Minor                 & \cite{2018ApJ...853..154A,2016JCAP...02..039M,2017ApJ...834..110A,2020JCAP...02..012H,2012APh....37...26M,2014PhRvD..89d2001A,2015PhRvL.115w1301A,2010ApJ...712..147A,2010JCAP...12..015D,2011PhRvL.107x1302A,2012PhRvD..86b1302G,2012JCAP...11..048H,2015ApJS..219...37P,2016PhRvD..93h3513Z,2016PhRvD..93j3009B,2019JCAP...04..048Q,2018JCAP...03..010P,2016MNRAS.461.3976P,2018JCAP...10..029C,2020JCAP...09..004A,2019JCAP...04..048Q,2016PhRvD..94j3502L,2015JCAP...09..016H,2012PhRvD..86b3528C,2011PhRvL.107x1303G,2018PhRvD..97h3007L} \\
32 & Willman 1                  & \cite{2016JCAP...02..039M,2017ApJ...834..110A,2014PhRvD..89d2001A,2015PhRvL.115w1301A,2010ApJ...712..147A,2015ApJS..219...37P,2016PhRvD..93h3513Z,2019JCAP...04..048Q,2018JCAP...03..010P,2016MNRAS.461.3976P,2021PhRvD.104h3037L,2018JCAP...10..029C,2019JCAP...04..048Q,2018PhRvD..97l2001L,2016PhRvD..94j3502L,2015JCAP...09..016H,2018PhRvD..97h3007L} \\  
 \bottomrule
\end{tabularx} 
\end{adjustwidth} 
 \end{table}

\begin{table}[H]\ContinuedFloat
\centering \small
\caption{{\em Cont.}} \label{tab:fermihawc-targetsref}  
	\begin{adjustwidth}{-\extralength}{0cm}
\newcolumntype{C}{>{\centering\arraybackslash}X}
\begin{tabularx}{\fulllength}{CCC}
\toprule

\textbf{Index in Figure \ref{fig:skymap_fermihawc}}	& \textbf{Source Name}	&  \textbf{Reference}\\
\midrule
\multicolumn{3}{c}{Dwarf spheroidal galaxy candidates}\\\midrule
33 & Canis Major                & \cite{2014PhRvD..89d2001A,2015PhRvL.115w1301A,2015ApJS..219...37P,2016PhRvD..94j3502L,2015JCAP...09..016H,2018PhRvD..97h3007L} \\
34 & Cetus II                   & \cite{2016PhRvD..93d3518L,2017ApJ...834..110A,2021PhRvD.104h3037L,2018PhRvD..97l2001L,2016PhRvD..94j3502L,2018PhRvD..97h3007L} \\
35 & Columba I                  & \cite{2016PhRvD..93d3518L,2017ApJ...834..110A,2016PhRvD..94j3502L,2018PhRvD..97h3007L} \\
36 & Eridanus II                & \cite{2017ApJ...834..110A,2016PhRvD..94j3502L,2015JCAP...09..016H,2015ApJ...809L...4D,2018PhRvD..97h3007L} \\
37 & Eridanus III               & \cite{2017ApJ...834..110A,2016PhRvD..94j3502L,2015JCAP...09..016H,2015ApJ...809L...4D,2018PhRvD..97h3007L} \\
38 & Grus I                     & \cite{2017ApJ...834..110A,2018JCAP...10..029C,2020JCAP...02..012H,2016PhRvD..94j3502L,2015JCAP...09..016H,2018PhRvD..97h3007L} \\
39 & Grus II                    & \cite{2016PhRvD..93d3518L,2017ApJ...834..110A,2016PhRvD..94j3502L,2018PhRvD..97h3007L} \\
40 & Horologium II              & \cite{2017ApJ...834..110A,2016PhRvD..94j3502L,2018PhRvD..97h3007L} \\
41 & Indus I/Kim 2              & \cite{2015JCAP...09..016H,2015ApJ...809L...4D,2017ApJ...834..110A,2018PhRvD..97h3007L} \\
42 & Indus II                   & \cite{2016PhRvD..93d3518L,2017ApJ...834..110A,2016PhRvD..94j3502L,2018PhRvD..97h3007L} \\
43 & Pegasus III                & \cite{2017ApJ...834..110A,2020JCAP...02..012H,2016PhRvD..94j3502L,2018PhRvD..97h3007L} \\
44 & Phoenix II                 & \cite{2017ApJ...834..110A,2016PhRvD..94j3502L,2015JCAP...09..016H,2015ApJ...809L...4D,2018PhRvD..97h3007L} \\
45 & Pictor I                   & \cite{2017ApJ...834..110A,2016PhRvD..94j3502L,2015JCAP...09..016H,2015ApJ...809L...4D,2018PhRvD..97h3007L} \\
46 & Reticulum III              & \cite{2016PhRvD..93d3518L,2017ApJ...834..110A,2016PhRvD..94j3502L,2018PhRvD..97h3007L} \\
47 & Sagittarius II             & \cite{2017ApJ...834..110A,2016PhRvD..94j3502L,2018PhRvD..97h3007L} \\
48 & Triangulum II              & \cite{2020PDU....2800529A,2020PhRvD.101j3001A,2018ApJ...853..154A,2017ApJ...834..110A,2017JCAP...11..003B,2021PhRvD.104h3037L,2018PhRvD..97l2001L,2016PhRvD..94j3502L,2015JCAP...09..016H,2018PhRvD..97h3007L} \\
49 & Tucana III                 & \cite{2016PhRvD..93d3518L,2017ApJ...834..110A,2021PhRvD.104h3037L,2018PhRvD..97l2001L,2016PhRvD..94j3502L,2018PhRvD..97h3007L} \\
50 & Tucana IV                  & \cite{2016PhRvD..93d3518L,2017ApJ...834..110A,2021PhRvD.104h3037L,2018PhRvD..97l2001L,2016PhRvD..94j3502L,2018PhRvD..97h3007L} \\
51 & Tucana V                   & \cite{2016PhRvD..93d3518L,2017ApJ...834..110A,2016PhRvD..94j3502L,2018PhRvD..97h3007L} \\ \toprule
\multicolumn{3}{c}{dwarf Irregular galaxies/irregular \& low surface-brightness objects}\\\midrule
52& Andromeda IV               & \cite{2019ICRC...36..520H} \\
53& DDO 43                     & \cite{2019ICRC...36..520H} \\
54& DDO 52                     & \cite{2019ICRC...36..520H} \\
55& DDO 101                    & \cite{2019ICRC...36..520H} \\
56& DDO 125                    & \cite{2019ICRC...36..520H} \\
57& DDO 133                    & \cite{2019ICRC...36..520H} \\
58& DDO 154                    & \cite{2019ICRC...36..520H} \\
59& DDO 168                    & \cite{2019ICRC...36..520H} \\
60& DDO 210/Aquarius           & \cite{2021PhRvD.104h3026G} \\
61& DDO 216/Pegasus dIrr       & \cite{2021PhRvD.104h3026G} \\
62& Haro 29                    & \cite{2019ICRC...36..520H} \\
63& Haro 36                    & \cite{2019ICRC...36..520H} \\
64& IC 10                      & \cite{2019ICRC...36..520H,2021PhRvD.104h3026G} \\
65& IC 1613                    & \cite{2019ICRC...36..520H,2018PhRvL.120j1101L,2021PhRvD.104h3026G} \\
66& Large Magellanic Cloud     & \cite{2015PhRvD..91j2001B} \\
67& NGC 3741                   & \cite{2019ICRC...36..520H} \\
68& NGC 6822                   & \cite{2021PhRvD.104h3026G} \\
69& Phoenix                    & \cite{2021PhRvD.104h3026G} \\
70& Small~Magellanic~Cloud     & \cite{2016PhRvD..93f2004C} \\
71& Smith High-Velocity Cloud  & \cite{2014ApJ...790...24D} \\
72& UGC 11583                  & \cite{2019ICRC...36..520H} \\
73& UGC 1281                   & \cite{2019ICRC...36..520H} \\
74& UGC 1501                   & \cite{2019ICRC...36..520H} \\
75& UGC 2455                   & \cite{2019ICRC...36..520H} \\
76& UGC 3371                   & \cite{2021MNRAS.501.4238B} \\
77& UGC 5272                   & \cite{2019ICRC...36..520H} \\
78& UGC 5427                   & \cite{2019ICRC...36..520H} \\
79& UGC 5918                   & \cite{2019ICRC...36..520H} \\
80& UGC 7047                   & \cite{2019ICRC...36..520H} \\
81& UGC 7232                   & \cite{2019ICRC...36..520H} \\
82& UGC 7559                   & \cite{2019ICRC...36..520H} \\
83& UGC 7603                   & \cite{2019ICRC...36..520H} \\
 \bottomrule
\end{tabularx} 
\end{adjustwidth} 
 \end{table}

\begin{table}[H]\ContinuedFloat
\centering \small
\caption{{\em Cont.}} \label{tab:fermihawc-targetsref}  
	\begin{adjustwidth}{-\extralength}{0cm}
\newcolumntype{C}{>{\centering\arraybackslash}X}
\begin{tabularx}{\fulllength}{CCC}
\toprule

\textbf{Index in Figure \ref{fig:skymap_fermihawc}}	& \textbf{Source Name}	&  \textbf{Reference}\\
\midrule
84& UGC 7861                   & \cite{2019ICRC...36..520H} \\
85& UGC 7866                   & \cite{2019ICRC...36..520H} \\
86& UGC 7916                   & \cite{2019ICRC...36..520H} \\
87& UGC 8508                   & \cite{2019ICRC...36..520H} \\
88& UGC 8837                   & \cite{2019ICRC...36..520H} \\
89& WLM                        & \cite{2019ICRC...36..520H,2021PhRvD.104h3026G} \\\toprule
\multicolumn{3}{c}{Spiral galaxies}\\\midrule
90 & M 31                       & \cite{2010JCAP...12..015D,2019PhRvD..99l3027D, 2021PhRvD.103b3027K,2016JCAP...12..028L,2018JCAP...06..043A} \\
91 & M 33                       & \cite{2019PhRvD..99l3027D} \\
92 & UGC 11707                  & \cite{2021MNRAS.501.4238B} \\
93 & UGC 12632                  & \cite{2021MNRAS.501.4238B} \\
94 & UGC 12732                  & \cite{2021MNRAS.501.4238B} \\\toprule
\multicolumn{3}{c}{Galaxy clusters (see caption)}\\\midrule
95 & 3C 129                     & \cite{2013ApJ...762L..22H,2016JCAP...02..026A,2016PhRvD..93j3525L} \\
96& A 1060/Hydra I             & \cite{2012JCAP...01..042H,2013ApJ...762L..22H,2016JCAP...02..026A,2016PhRvD..93j3525L,2018PhRvL.120j1101L} \\
97& A 1367                     & \cite{2012JCAP...01..042H,2012JCAP...07..017A,2013ApJ...762L..22H,2016JCAP...02..026A,2016PhRvD..93j3525L} \\
98& A 2877                     & \cite{2013ApJ...762L..22H,2016JCAP...02..026A,2016PhRvD..93j3525L} \\
99& A 3627/Norma               & \cite{2013ApJ...762L..22H,2016JCAP...02..026A,2016PhRvD..93j3525L,2018PhRvL.120j1101L} \\
100 & ACO S 636                  & \cite{2012JCAP...01..042H,2013ApJ...762L..22H,2016JCAP...02..026A,2016PhRvD..93j3525L} \\
101 & AWM 7                      & \cite{2010JCAP...12..015D,2012JCAP...01..042H,2012JCAP...11..048H,2010JCAP...05..025A,2013ApJ...762L..22H, 2016JCAP...02..026A,2016PhRvD..93j3525L} \\
102 & {Centaurus   (A 3526)}         & \cite{2010JCAP...12..015D,2012JCAP...07..017A,2012JCAP...11..048H,2010JCAP...05..025A,2013ApJ...762L..22H, 2016JCAP...02..026A,2021MNRAS.502.4039T,2018PhRvL.120j1101L} \\
103 & Coma cluster               & \cite{2010JCAP...12..015D,2010JCAP...05..025A,2012JCAP...01..042H,2012JCAP...07..017A,2012JCAP...11..048H,2012MNRAS.427.1651H,2013ApJ...762L..22H,2016JCAP...02..026A,2016PhRvD..93j3525L,2021MNRAS.502.4039T} \\
104 & Fornax cluster             & \cite{2010JCAP...05..025A,2010JCAP...12..015D,2012JCAP...01..042H,2012JCAP...07..017A,2012MNRAS.427.1651H,2012JCAP...11..048H,2013ApJ...762L..22H,2016JCAP...02..026A,2016PhRvD..93j3525L,2021MNRAS.502.4039T,2018PhRvL.120j1101L} \\
105 & M 49                       & \cite{2010JCAP...12..015D,2012JCAP...07..017A,2012JCAP...11..048H,2010JCAP...05..025A,2013ApJ...762L..22H,2016JCAP...02..026A,2016PhRvD..93j3525L} \\
106 & NGC 4636                   & \cite{2010JCAP...12..015D,2012JCAP...01..042H,2012JCAP...07..017A,2012JCAP...11..048H,2010JCAP...05..025A, 2016JCAP...02..026A,2013ApJ...762L..22H,2016PhRvD..93j3525L} \\
107 & NGC 5044                   & \cite{2012JCAP...07..017A,2013ApJ...762L..22H} \\
108 & NGC 5813                   & \cite{2012JCAP...01..042H,2012JCAP...07..017A,2013ApJ...762L..22H,2016JCAP...02..026A,2016PhRvD..93j3525L} \\
109 & NGC 5846                   & \cite{2012JCAP...07..017A,2013ApJ...762L..22H} \\
110 & Ophiuchus                  & \cite{2013ApJ...762L..22H,2016JCAP...02..026A,2016PhRvD..93j3525L} \\
111 & Perseus                    & \cite{2013ApJ...762L..22H,2016JCAP...02..026A,2021MNRAS.502.4039T,2016PhRvD..93j3525L} \\
112 & Virgo cluster              & \cite{2012PhRvD..86g6004M,2012JCAP...11..048H,2013ApJ...762L..22H,2015ApJ...812..159A, 2016JCAP...02..026A,2021MNRAS.502.4039T,2017AIPC.1792f0010C,2018PhRvL.120j1101L,2012MNRAS.427.1651H,2016PhRvD..93j3525L} \\
\bottomrule
\end{tabularx}
\end{adjustwidth} 
\end{table}
	
\normalsize
\begin{adjustwidth}{-\extralength}{0cm}
\printendnotes[custom]
\end{adjustwidth}

\begin{adjustwidth}{-\extralength}{0cm}

\reftitle{References}




\end{adjustwidth}
\end{document}